\documentstyle[aps,epsf]{revtex}

\begin{document}

\thispagestyle{empty}

\hbox to \hsize{\hfill
\vbox{\bf 
\hbox{FERMILAB-Pub-99-187-T}
\hbox{MADPH-99-1122}
\hbox{AMES-HET-99-05}
\hbox{\bf hep-ph/9906487}}}

\begin{center}
\vspace{0.7 cm}
\large
{\bf Long Baseline Neutrino Physics with a Muon Storage Ring Neutrino Source}

\vspace{1.0cm}

V. Barger$^1$, S. Geer$^2$, and K. Whisnant$^3$
\vspace{0.5cm}

\it
$^1$Department of Physics, University of Wisconsin,
Madison, WI 53706, USA\\
$^2$Fermi National Accelerator Laboratory, P.O. Box 500,
Batavia, IL 60510, USA\\
$^3$Department of Physics and Astronomy, Iowa State University,
Ames, IA 50011, USA
\end{center}

\vspace{1cm}

\begin{abstract}
We examine the physics capabilities of known flavor neutrino
beams from intense muon sources. We find that long-baseline neutrino
experiments based on such beams can provide precise measurements of
neutrino oscillation mass and mixing parameters. Furthermore, they can
test whether the dominant atmospheric neutrino oscillations are $\nu_\mu
\rightarrow \nu_\tau$ and/or $\nu_\mu \rightarrow \nu_s$, determine the
$\nu_\mu \rightarrow \nu_e$ content of atmospheric neutrino
oscillations, and measure $\nu_e \rightarrow \nu_\tau$ appearance.
Depending on the oscillation parameters, they may be able to detect
Earth matter and $CP$ violation effects and to determine the ordering of
some of the mass eigenstates.
\end{abstract}

\vspace{0.5cm}
14.60.Pq, 13.15.+g, 13.35.Bv, 07.77Ka

\vspace{0.5cm}

\section{Introduction}

In recent years solar-neutrino and atmospheric-neutrino measurements
have provided a growing body of evidence for the existence of neutrino
oscillations~\cite{review}.  The observed solar neutrino
deficit~\cite{solar,solarth} can be interpreted as evidence for the
oscillation of electron neutrinos ($\nu_e$) into neutrinos of a
different flavor. The recent atmospheric-neutrino results from the
Super-Kamiokande collaboration~\cite{superk}, along with other
experiments~\cite{atmos} suggest the oscillation of muon neutrinos
($\nu_\mu$) into neutrinos of a different flavor, which are
dominantly either tau neutrinos ($\nu_\tau$) or sterile neutrinos
($\nu_s$). Taken together, these results suggest the mixing of at least
three different neutrino types~\cite{three,bww98,global,hall2}. In addition,
the LSND collaboration has reported~\cite{lsnd1} results from $\mu^+$
decays at rest that could be interpreted as the first evidence for
$\overline{\nu}_\mu \rightarrow \overline{\nu}_e$ oscillations in an
accelerator based experiment. LSND has also reported~\cite{lsnd2}
results from measurements of $\pi^+$ decays in flight that could be the
first evidence for $\nu_\mu \rightarrow \nu_e$ oscillations. If all of
the above reported effects survive, then oscillations to a sterile
neutrino are required, since there would then be three distinct
mass-squared difference ($\Delta m^2$) values~\cite{four}.

The solar-neutrino, atmospheric-neutrino, and LSND results have
generated much interest in future accelerator-based neutrino
oscillation experiments. We can anticipate that an extensive
experimental program will be needed to firmly establish the existence of
neutrino oscillations and to precisely determine all the parameters
relevant to the phenomenon. The up-coming accelerator-based
experiments~\cite{K2K,MINOS} will firmly ground the existence of
neutrino oscillations, and measure some of the associated parameters.
Also, reactor experiments will tightly constrain $\nu_e$
disappearance~\cite{chooz,bugey,kam,paloverde}; a sensitivity down to
$10^{-5}$~eV$^2$ is expected in Kamland.
However, precise measurements of neutrino mass-squared differences
and the mixing matrix that relates neutrino-flavor eigenstates to
neutrino-mass eigenstates will almost certainly require a further
generation of experiments exploiting higher intensity and/or higher
quality neutrino beams than currently
available~\cite{sg_prd,derujula,camp}. Other goals of such experiments
might be (i) to determine whether the atmospheric neutrino oscillations
are $\nu_\mu \rightarrow \nu_\tau$, $\nu_\mu \rightarrow \nu_s$, or a
mixture of both~\cite{matter,raby}, (ii) to detect Earth matter effects on
neutrino oscillations~\cite{wolf,bppw}, (iii) when matter effects
are present, to determine the ordering of the neutrino mass eigenstates
responsible for the oscillation being measured~\cite{bppw,lang,lipari2},
and (iv) to detect $CP$ and $T$ violation if it exists in the lepton
sector~\cite{derujula,cabbibo,bpw,pakvasa,CP}.

It has been suggested~\cite{sg_prd} that 
higher intensity and higher quality neutrino beams could be 
made by exploiting the very intense muon sources that are 
currently being developed as a part of ongoing muon 
collider~\cite{status_report} feasibility studies. 
The muons would be accelerated up to the desired 
energy, and injected into a storage ring consisting of two long 
straight sections joined together by two arcs. Muons that decay in 
the straight sections would form neutrino beams consisting of 50\% 
$\nu_\mu$ and 50\% $\overline{\nu}_e$ if negative muons are stored, 
and 50\% $\nu_e$ and 50\% $\overline{\nu}_\mu$ if positive muons 
are stored. A compact muon storage ring neutrino source could be 
tilted downwards at a large angle to enable neutrino beams 
to be directed through the Earth. 
If the muons from a muon collider muon source are accelerated 
to energies of $\ge 10$~GeV and injected into a suitable 
storage ring, it has been shown that the 
neutrino fluxes are sufficient to detect hundreds of neutrino 
charged current (CC) interactions per year in a reasonable size 
detector on the other side of the Earth~\cite{sg_prd}. 

The muon storage ring neutrino source idea has led to several recent
workshops~\cite{ws1,ws2}. A number of
papers~\cite{sg_prd,derujula,bww,papers} have discussed the physics
potential of this new type of neutrino facility.  In addition, a
preliminary muon storage ring neutrino source design study has been
made~\cite{design}, and further studies are planned. The evolution of
the existing accelerator complexes at
Fermilab~\cite{summer_study,vision} and CERN~\cite{cern_vision} towards
a muon collider with a muon storage ring neutrino source has also been
considered. Much further work will be required to develop a realistic
design before the first muon storage ring neutrino source can be
proposed.

In this paper we consider the physics potential of muon storage ring
neutrino sources that are being considered in explicit upgrade scenarios
for the Fermilab accelerator complex. Two different geometries are
considered: (i) a neutrino source at Fermilab pointing toward the Soudan
mine in Minnesota ($L = 732$~km), and (ii) a neutrino source at Fermilab
pointing toward the Gran Sasso underground laboratory in Italy (baseline
length $L = 7332$~km).

The paper is organized as follows. In Section~II the characteristics of
neutrino beams from muon storage ring sources are discussed and the
basic oscillation formulas are presented.  The role of muon storage ring
neutrino sources in exploring the parameters associated with $\nu_\mu
\rightarrow \nu_\tau$ and $\nu_\mu \rightarrow \nu_s$ oscillations is
considered in Section~III.  Section~IV discusses $\nu_e \rightarrow
\nu_\mu$ oscillations, including matter effects and the possibility of
detecting $CP$ violation. In Section~V we discuss $\nu_e \rightarrow
\nu_\tau$ oscillations.  Finally, a summary is given in Section~VI.

\section{Neutrino beams from a Muon Storage Ring}

The neutrino fluxes and interaction rates at a given location downstream
of a muon storage ring neutrino source depend upon the number of stored
muons, the beam divergence within the neutrino-beam forming straight
section, and the energies and polarization of the decaying muons. In
this section we describe the storage ring parameters, discuss the muon
decay kinematics, present the calculated neutrino fluxes and interaction
rates at $L = 732$~km and 7332~km, and discuss the matter effects for
$\nu_e \rightarrow \nu_\mu$, $\nu_e \rightarrow \nu_\tau$, and $\nu_\mu
\rightarrow \nu_s$ oscillations.

\subsection{Storage ring parameters}

A preliminary design study for a muon storage ring neutrino 
source at Fermilab is described in Ref.~\cite{design}. 
The neutrino source consists of:
\begin{description}
\item{(i)} An upgraded high-intensity proton source~\cite{summer_study} 
that cycles at 15~Hz, and delivers 12 bunches per cycle, each 
containing $2.5 \times 10^{12}$ protons at 16~GeV.
\item{(ii)} A pion production target, pion collection system, 
and decay channel. The 16~GeV protons interact in the target 
to produce per incident proton approximately 0.6 charged pions 
of each sign captured within the decay channel. At the end of 
a 50~m long decay channel on average 0.2 muons of each charge 
are produced for each proton incident on the production target. 
Hence there would be about $6 \times 10^{12}$ muons of the 
desired charge at the end of the decay channel per accelerator 
cycle, and therefore $9 \times 10^{20}$ muons per operational 
year ($10^{7}$~secs). 
\item{(iii)} A muon cooling channel that captures the muons 
exiting the decay channel into bunches with rms lengths 
$\sigma_z = 1.5$~m, and a normalized transverse emittance 
$\epsilon_N \sim 0.017$~m-rad. The cooling channel reduces the 
emittance to $\epsilon_N \sim 0.005$~m-rad. At the end of the 
cooling channel there would be about $5.4 \times 10^{12}$ muons 
of the desired charge per accelerator cycle, contained within 
12 bunches with rms lengths $\sigma_z = 2$~m, a mean muon energy 
of 230~MeV, and an energy spread given by $\sigma_E / E \sim 0.2$. 
Hence, there would be $8.1 \times 10^{20}$ cooled muons per 
operational year. 
\item{(iv)} An acceleration system that captures the muons from 
each long bunch exiting the cooling channel into 16 short bunches 
with $\sigma_z \sim 0.9$~cm, 
and accelerates the muons to 10~GeV. About 60\% of the muons 
survive the capture and acceleration stage. Hence, about 
$5 \times 10^{20}$ muons are accelerated to 10~GeV per 
operational year.
\item{(v)} A muon storage ring with two long straight sections. 
The circumference of the ring is 448~m, and the length of the 
neutrino beam-forming straight section is 150~m. Hence about 
33\% of the muon decays contribute to the neutrino beam, 
and there are $1.6 \times 10^{20}$ neutrinos and $1.6 \times 10^{20}$ 
antineutrinos in the beam per operational year. 
\end{description}

Note that in the scenario described above the proton source provides 
approximately one-third of the beam power of the source required for 
a muon collider. The fluxes and interaction rates discussed 
in the following sections should be multiplied by a factor of 3.3 to 
obtain the results that would correspond to a muon storage ring neutrino 
source that utilizes the full muon collider muon source.

\subsection{Fluxes and Interaction Rates}

In the muon rest-frame the distribution of muon 
antineutrinos (neutrinos) from the decay 
$\mu^{\pm} \rightarrow  e^{\pm}$ + $\nu_e$ ($\overline{\nu}_e)$ + 
$\overline{\nu}_{\mu}$ ($\nu_{\mu}$) of polarized muons is given by 
the expression~\cite{gaisser}
\begin{equation}
\frac{d^2N_{\nu_{\mu}}}{dxd\Omega} = \frac{2x^2}{4\pi}
\left[ (3-2x) \mp (1-2x)\cos\theta \right] ,
\label{eq0}
\end{equation}
where $x \equiv 2E_\nu$/$m_\mu$, $\theta$ is the angle between 
the neutrino momentum vector and the muon spin direction, and $m_\mu$ is the 
muon rest mass. The corresponding expression describing the 
distribution of electron neutrinos (antineutrinos) is
\begin{equation}
\frac{d^2N_{\nu_{e}}}{dxd\Omega} = \frac{12x^2}{4\pi}
\left[ (1-x) \mp (1-x)\cos\theta \right] .
\end{equation}
Thus, the neutrino and antineutrino energy- and angular-
distributions depend upon the parent muon 
energy, the decay angle, and the direction of the 
muon spin vector. 
For an ensemble of muons we must average over the polarization 
of the initial state muons, and the distributions become
\begin{equation}
\frac{d^2N_{\nu_{\mu}}}{dxd\Omega} \propto \frac{2x^2}{4\pi}
\left[ (3-2x) \mp (1-2x)P_{\mu}\cos\theta \right] ,
\label{eq1}
\end{equation}
and 
\begin{equation}
\frac{d^2N_{\nu_{e}}}{dxd\Omega} \propto \frac{12x^2}{4\pi}
\left[ (1-x) \mp (1-x)P_{\mu}\cos\theta \right] \; ,
\label{eq2}
\end{equation}
where $P_{\mu}$ is the average muon polarization along the 
chosen quantization axis, which in this case is the 
beam direction.

\begin{figure}
\epsfxsize2.25in
\centerline{\epsffile{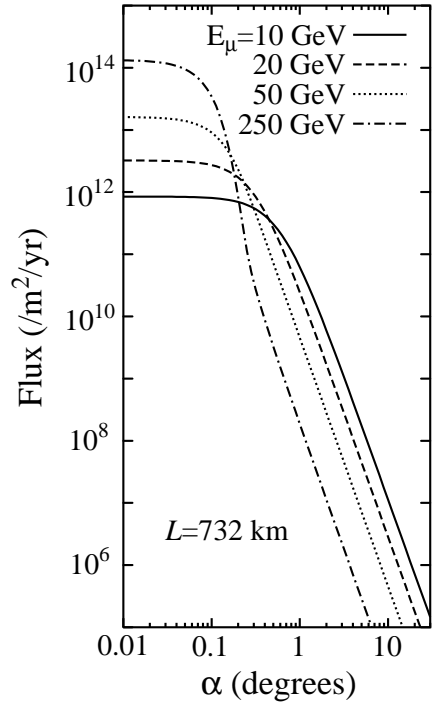}}
\vspace{.25cm}
\caption{Total neutrino flux of a given flavor versus angle at the
source from the beamline at
$L=732$~km for $E_\mu = 10$~GeV (solid line), 20~GeV (dashed), 50~GeV
(dotted), and 250~GeV (dot-dashed), assuming $1.6\times10^{20}$ neutrinos
in the beam per operational year. A Gaussian muon beam divergence of
1~mr has been folded in. The total antineutrino flux is the same.
}
\label{flux1_fig}
\end{figure}

\begin{figure}
\epsfxsize6in
\centerline{\epsffile{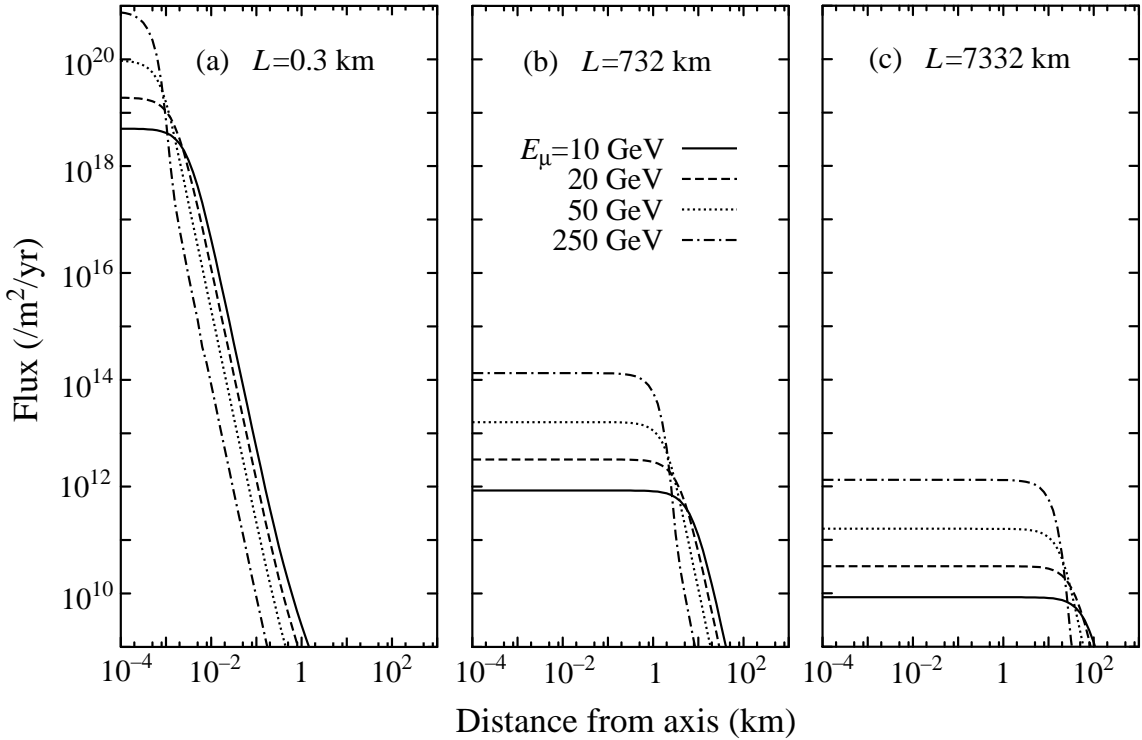}}
\vspace{.25cm}
\caption{Total neutrino flux of a given flavor versus perpendicular
distance from the beam axis at (a) $L=0.3$~km, (b) $732$~km, and (c)
$7332$~km (dotted) for for $E_\mu =$~10~GeV (solid line), 20~GeV
(dashed), 50~GeV (dotted), and 250~GeV (dot-dashed), assuming
$1.6\times10^{20}$ neutrinos in the beam per operational year. A
Gaussian muon beam divergence of 1~mr has been folded in. The total
antineutrino flux is the same.
}
\label{flux2_fig}
\end{figure}

Using Eqs.~\ref{eq1} and \ref{eq2}, the calculated 
$\nu_\mu$, $\nu_e$, $\overline{\nu}_\mu$, and $\overline{\nu}_e$ 
fluxes at a muon storage ring neutrino source 
were presented in Ref.~\cite{sg_prd} 
as a function of L, $P_{\mu}$, and the energy of the stored muons 
($E_\mu$). For unpolarized muons, these results can be summarized
by the formula
\begin{equation}
\Phi \equiv {d^2 N_\nu\over dA dt} =
{n_0\over 4\pi L^2 \gamma^2(1-\beta\cos\alpha)^2} \; ,
\end{equation} 
where $n_0$ is the number of neutrinos (or antineutrinos) per unit
time in a given beam, $\gamma=E_\mu/m_\mu$, $\beta=p_\mu/E_\mu$,
$\alpha$ is the angle between the beam axis and the direction of
interest, and $dA$ is the differential area at the detector. In
practice the muon beam within the storage ring will have a
finite divergence which must be taken into account. Current design
studies suggest a typical divergence of $\sim$~1~mr within the
beam-forming straight section. The neutrino flux at $L=732$~km
is shown versus $\alpha$ in Fig.~\ref{flux1_fig} for several muon
energies. In Fig.~\ref{flux2_fig} the neutrino flux is shown versus
the perpendicular distance from the beam axis for several values of
$L$ and $E_\mu$. As Fig.~\ref{flux2_fig} shows, the flux is fairly
uniform near the axis, and, provided $1/\gamma > 1$~mr, begins to
fall off only for distances of order $L/\gamma$ away from the beam.
In the extreme forward direction $\cos\alpha \simeq 1$, and using
the approximation $\beta \simeq 1 -  1/(2\gamma^2)$, the neutrino
flux may be approximated as
\begin{equation}
\Phi \simeq {n_0 \gamma^2 \over\pi L^2} \; .
\end{equation}
These calculated fluxes are summarized in Table~\ref{flux_tab} for 
the baseline lengths $L = 732$~km and $L = 7332$~km, and 
the storage ring parameters described Section~II.A.

\begin{figure}
\epsfxsize4.5in
\centerline{\epsffile{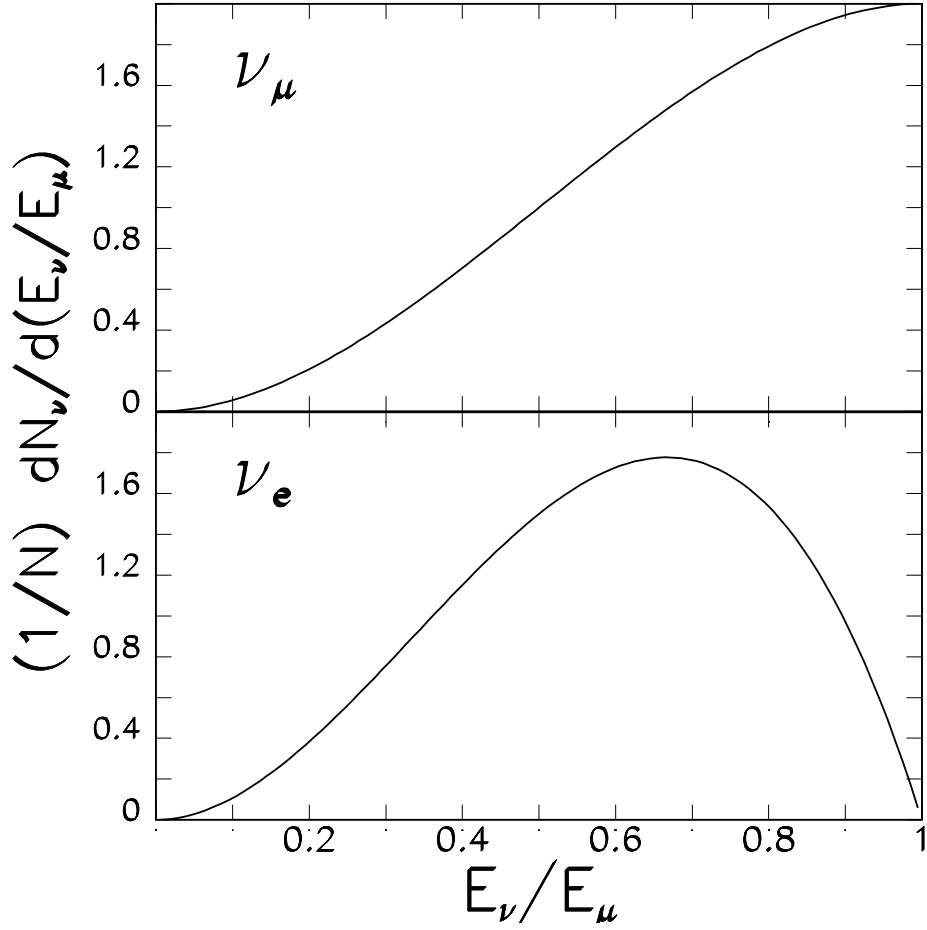}}
\vspace{-1cm}
\caption{Neutrino and antineutrino spectra for muon-type neutrinos 
(top) and electron-type neutrinos (bottom) in the beam downstream 
of a muon storage ring neutrino source containing unpolarized muons. 
}
\label{enu_fig}
\end{figure}

Equations~\ref{eq1} and \ref{eq2} can also be used to calculate 
the distribution $dN_\nu/dE_\nu$ of $\nu_\mu$ and $\nu_e$ energies 
in the beam  using the approximation that, in the laboratory frame, 
$x \rightarrow E_\nu / E_\mu$ for forward neutrinos from high energy
muon decays ($E_\mu >> m_\mu$). 
The calculated $\nu_\mu$ and $\nu_e$ spectra are shown in 
Fig.~\ref{enu_fig}. Note that the $\nu_\mu$ 
spectrum peaks at the stored muon beam energy, whereas the $\nu_e$ 
spectrum peaks at two-thirds of the stored muon beam energy. 
Also, since $\Phi \propto E_\mu^2$ and $\sigma \propto E_\mu$, the
total event rate is proportional to $E_\mu^3$.

\begin{figure}
\epsfxsize4.5in
\centerline{\epsffile{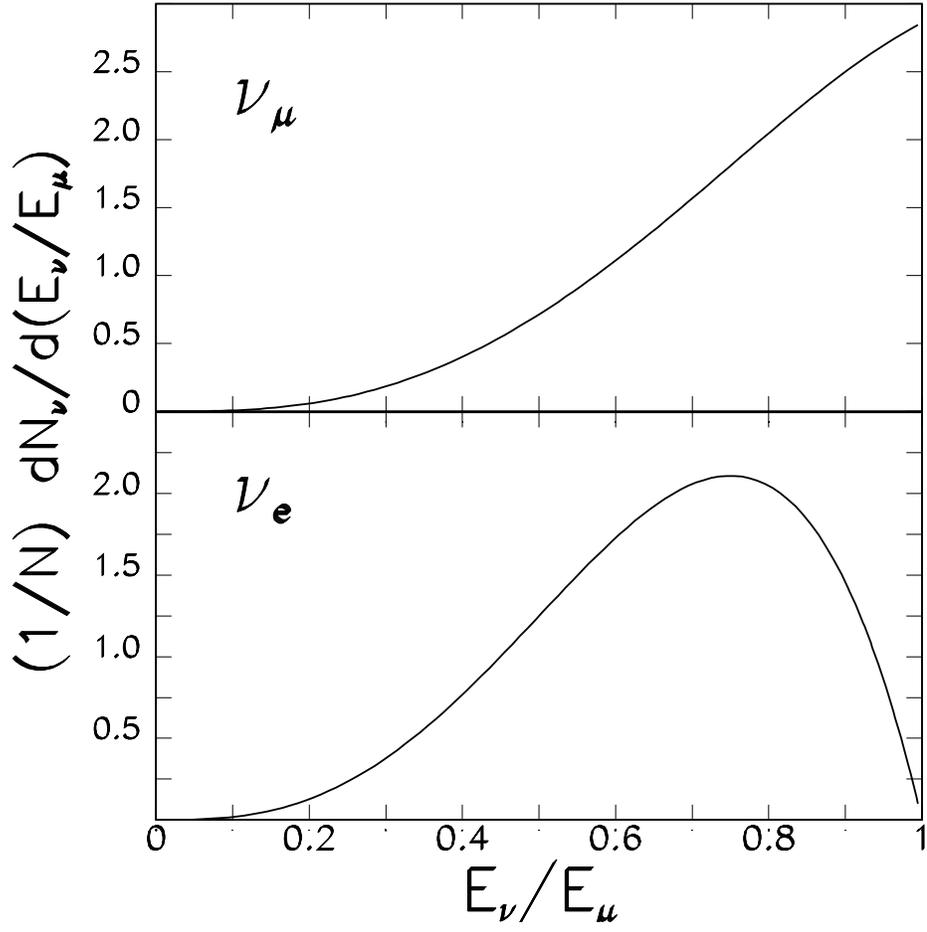}}
\vspace{-1cm}
\caption{Event energy distributions for the CC interactions of 
muon-type neutrinos (top) and electron-type neutrinos (bottom) 
downstream
of a muon storage ring neutrino source containing unpolarized muons.
}
\label{ecc_fig}
\end{figure}

The CC neutrino and antineutrino interaction rates 
can be calculated using the approximate expressions for 
the cross sections
\begin{equation}
\sigma_{\nu N} \;\sim\; 0.67 \times 10^{-38}\;{\rm cm}^2 \times E_\nu 
\;{\rm (GeV)} \;
\label{eq3}
\end{equation}
and
\begin{equation}
\sigma_{\overline{\nu} N} \;\sim\; 0.34 \times 10^{-38}\;{\rm cm}^2 
\times E_{\overline{\nu}} \;{\rm (GeV)} \; .
\label{eq4}
\end{equation}
The modification in the linear energy dependence due to the $W$
propagator can be neglected for the neutrino energies we consider. 
Using Eqs.~\ref{eq3} and \ref{eq4}, and assuming no neutrino
oscillations, the CC interaction rates corresponding to the fluxes in
Table~\ref{flux_tab} are shown in Table~\ref{rate_tab}. Given the
distribution of neutrino and antineutrino energies in the beam
(Fig.~\ref{enu_fig}), and the linear dependence of the CC cross sections
with energy (Eqs~\ref{eq3} and \ref{eq4}), the predicted energy
distributions for the resulting CC interactions are shown in
Fig.~\ref{ecc_fig}. Note that the $\nu_\mu$ CC event energies peak at
the stored muon beam energy, whereas the $\nu_e$ event energies peak at
about three-quarters of the stored muon beam energy. For $\nu_\tau$ and
$\bar\nu_\tau$ CC cross sections, we use the ratios
$\sigma_{CC}(\nu_\tau N)/\sigma_{CC}(\nu_\mu N)$ and
$\sigma_{CC}(\bar\nu_\tau N)/\sigma_{CC}(\bar\nu_\mu N)$
given in Ref.~\cite{goodman}, multiplied by the cross sections in
Eqs.~\ref{eq3} and \ref{eq4}. If the $\nu_\tau$ cross sections from
Ref.~\cite{casper} are used, the predicted event rates are 20--30\%
higher; we use the more conservative values of Ref.~\cite{goodman}.

Event rates for MINOS are compared to those for a Fermilab to Soudan
experiment using a muon storage ring in Table~\ref{compare_tab}. Since
for a muon storage ring neutrino source the total event rate increases
as $E_\mu^3$, increasing the muon energy dramatically improves the
usefulness of such a machine. From this comparison we conclude: (i) for
electron neutrino beams, even a very low energy muon storage ring might
be interesting, and (ii) for muon neutrino beams, rates become quite
large compared to conventional neutrino beams for $E_\mu \ge 20$~GeV.

\begin{figure}
\epsfxsize4.5in
\centerline{\epsffile{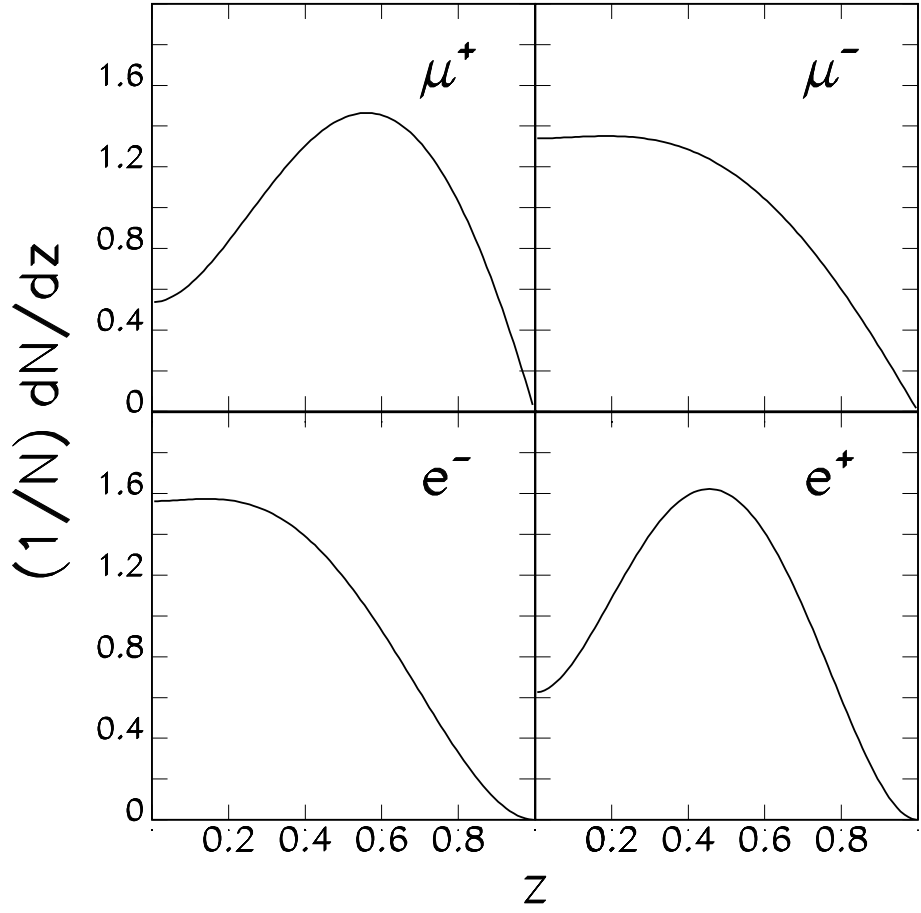}}
\vspace{-1.0cm}
\caption{Lepton energy spectra for CC $\overline{\nu}_\mu$ (top left),
$\nu_\mu$ (top right), $\nu_e$ (bottom left), and $\overline{\nu}_e$ 
(bottom right) interactions.}
\label{elept_fig}
\end{figure}

We next consider the distribution of energies of the charged leptons 
produced in the neutrino and antineutrino CC interactions 
downstream of a muon storage ring neutrino source. 
It is useful to define the scaling variable
\begin{equation}
y \;\equiv\; 1 - E_l / E_\nu \; ,
\end{equation}
where $E_l$ is the energy of a charged lepton produced 
the CC interaction. The differential neutrino and 
antineutrino cross sections are given approximately by
\begin{equation}
\frac{d\sigma_\nu}{dy} \;\propto \; E_\nu\;\left[1 
+ \frac{(1-y)^2}{5}\right] \; ,
\label{eq5}
\end{equation}
and
\begin{equation}
\frac{d\sigma_{\overline{\nu}}}{dy} \;\propto \; 
E_{\overline{\nu}}\;\left[ (1-y)^2 + \frac{1}{5} \right] \; .
\label{eq6}
\end{equation}
These formulas neglect the $Q^2$ dependence of the structure functions.
If {\it unpolarized} positive muons are stored in the muon storage ring
neutrino source, the $\overline{\nu}_\mu$ CC interaction rate  
in a distant detector is described by the doubly 
differential distribution obtained from Eqs.~\ref{eq0} 
and \ref{eq6}
\begin{equation}
\frac{d^2 N^{CC}_{\overline{\nu}_\mu}}{dxdy} \;\propto \;
x^3 \; (3 - 2x) \left[ (1-y)^2 + \frac{1}{5} \right] \; ,
\label{eq7}
\end{equation}
where, for a high energy muon storage ring, 
$x \rightarrow E_\nu / E_\mu$ 
in the laboratory frame. It is convenient to define 
the normalized charged lepton energy
\begin{equation}
z \;\equiv\; E_l / E_\mu \; .
\end{equation}
At high energies $(1 - y) \simeq z/x$.
Integrating Eq.~\ref{eq7} over $x$, the charged lepton 
energy spectrum is given by
\begin{equation}
\frac{dN_{\mu^+}}{dz} \;\propto\; 
\int_z^1 dx\; x^2 (3-2x) \left[ \left(\frac{z}{x} \right)^2 
+ \frac{1}{5} \right] \; ,
\end{equation}
yielding
\begin{equation}
\frac{dN_{\mu^+}}{dz} \;\propto\;
1\;+\;20z^2\;-\;32z^3\;+\;11z^4 \;.
\label{eq8}
\end{equation}
The corresponding expression for the electron energy spectrum 
arising from the CC $\nu_e$ interactions is
\begin{equation}
\frac{dN_{e^-}}{dz} \;\propto\;
5\;+\;6z^2\;-\;32z^3\;+\;21z^4 \;.
\label{eq9}
\end{equation}
The expressions for the $\mu^-$ and $e^+$ spectra resulting from 
CC interactions when unpolarized negative muons are stored in the muon storage 
ring neutrino source are respectively
\begin{equation}
\frac{dN_{\mu^-}}{dz} \;\propto\;
5\;+\;4z^2\;-\;16z^3\;+\;7z^4 \; ,
\label{eq10}
\end{equation}
and
\begin{equation}
\frac{dN_{e^+}}{dz} \;\propto\;
1\;+\;30z^2\;-\;64z^3\;+\;33z^4 \; .
\label{eq11}
\end{equation}
The charged lepton spectra computed from Eqs.~\ref{eq8}--\ref{eq11} 
are shown in Fig.~\ref{elept_fig}.

\subsection{Interaction Rates with Neutrino Oscillations}

\begin{figure}
\epsfxsize4.5in
\centerline{\epsffile{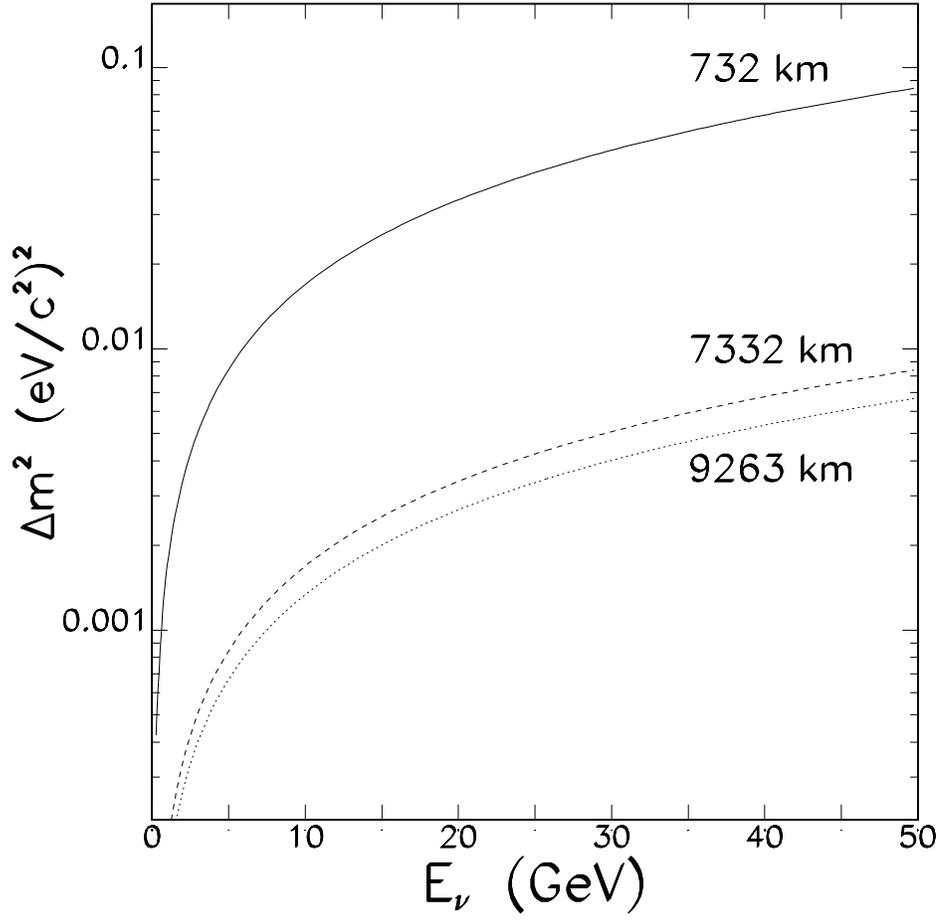}}
\vspace{-1.0cm}
\caption{The $\Delta m^2$ in Eq. 17 that yields a maximum vacuum
oscillation probability, shown versus the neutrino 
energy for three baseline lengths: (i) Fermilab $\rightarrow$ 
Soudan (solid line), (ii)  Fermilab $\rightarrow$ Gran Sasso 
(broken line), and (iii)  Fermilab $\rightarrow$ Kamioka Mine 
(dotted line).}
\label{dm2_e}
\end{figure}

\begin{figure}
\epsfxsize4.5in
\centerline{\epsffile{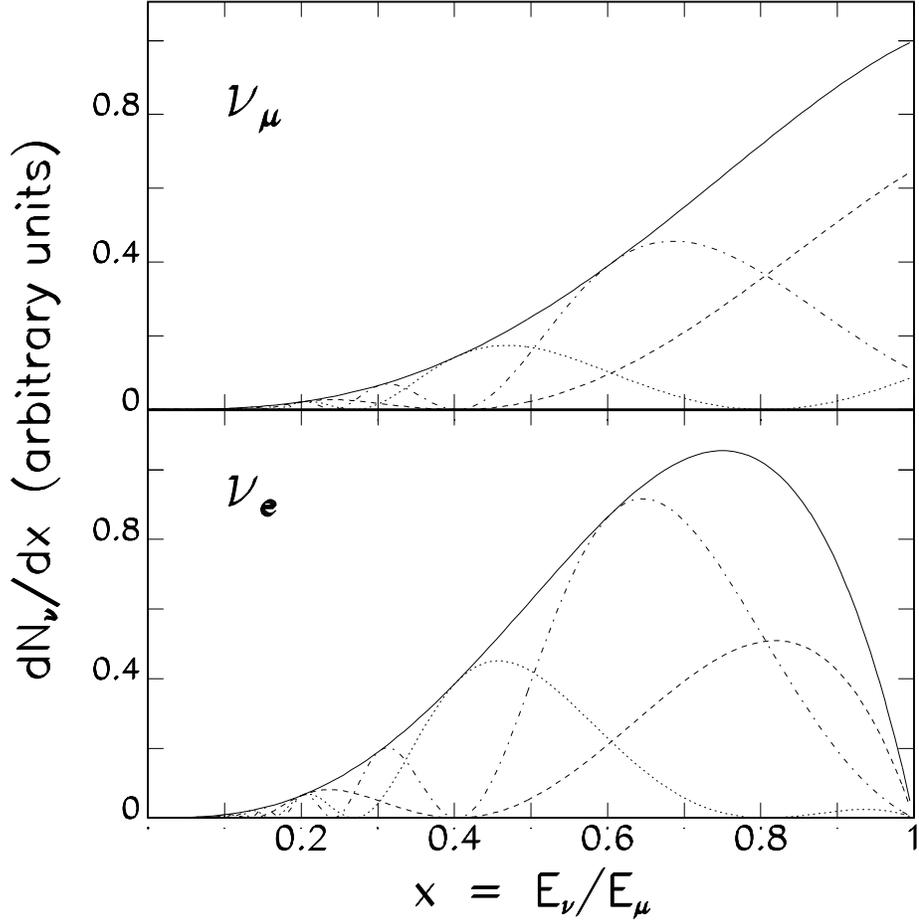}}
\vspace{-1.0cm}
\caption{Neutrino and antineutrino charged current interaction 
spectra for muon-type neutrinos 
(top) and electron-type neutrinos (bottom) downstream 
of a muon storage ring neutrino source containing unpolarized muons. 
The differential distributions are shown as a function of the parameter 
$\eta \equiv \Delta m^2 L/E_\mu$ for $\eta = 0.5$ (broken curves), 
$\eta = 1$ (dotted curves), and $\eta = 1.5$ (dot-dashed curves). 
The solid curves show the spectra in the absence of oscillations.
}
\label{enu_osc}
\end{figure}

For a given neutrino flavor, neutrino oscillations will modify the 
neutrino flux at a distant detector, 
and hence the associated charged current interaction rates. 
Within the framework of two-flavor vacuum oscillations, the  
flavor eigenstates $\nu_\alpha$ and $\nu_\beta$ 
are related to the mass eigenstates $\nu_i$ and $\nu_j$ 
by
\begin{equation}
\nu_\alpha \;=\; \nu_i \cos\theta \;-\; \nu_j \sin\theta \; , 
\end{equation}
\vspace{-0.3cm}
\begin{equation}
\nu_\beta \;=\; \nu_i \sin\theta \;+\; \nu_j \cos\theta \; ,
\end{equation}
where $\theta$ is the mixing angle. 
The probability that, while traversing a distance $L$ in vacuum, 
a neutrino of type $\alpha$ 
oscillates into a neutrino of type $\beta$ is given by
\begin{equation}
P(\nu_\alpha \rightarrow \nu_\beta) = 
\sin^2(2\theta) \; \sin^2(1.267\Delta m_{ji}^2 \; L/E_\nu) \; ,
\label{vprob}
\end{equation}
where $\Delta m_{ji}^2 \equiv m_j^2 - m_i^2$ 
is measured in eV$^2$/c$^4$, $L$ in km, and 
the neutrino energy $E_\nu$ is in GeV. 
The neutrino oscillation length in vacuum $L_V$ is given by
\begin{equation}
L_V \;=\; \frac{2.48\; E_\nu}{\Delta m^2_{ji}} \; .
\end{equation}
The first maximum in the oscillation probability occurs when 
$L = L_V/2$. 
The values of $\Delta m_{ji}^2$ that correspond to this oscillation 
maximum are shown in Fig.~\ref{dm2_e} as a function of $E_\nu$ for 
for three baseline lengths. Note that if $\Delta m_{ji}^2$ is small, 
short baseline lengths require low neutrino energies to probe the 
oscillation maximum. It is useful to define
\begin{equation}
\eta\;\equiv\;\frac{\Delta m_{ji}^2 L}{E_\mu}\;=\;
\frac{\Delta m_{ji}^2 L}{E_\nu}\;x \; .
\end{equation}
The first maximum in the oscillation probability occurs when 
$\eta = 1.24 x$. The modulation of the $\nu_e$ and $\nu_\mu$ 
CC interaction spectra for neutrinos originating from a muon storage
ring neutrino source is shown in Fig.~\ref{enu_osc} as a function of
$\eta$. The distributions of the $\nu_\alpha$ CC interaction energies
exhibit peaks whose locations are very sensitive to $\eta$ (and hence
$\Delta m_{ji}^2$) provided $\eta \sim 1$. 

If there are three neutrino types, then the flavor eigenstates are
related to the mass eigenstates by a $3\times3$ unitary matrix~\cite{MNS}
\begin{equation}
\left(\begin{array}{c} \nu_e \\ \nu_\mu \\ \nu_\tau \end{array} \\ \right)=
\left(\begin{array}{ccc}
c_{12}c_{13} & s_{12}c_{13} & s_{13}e^{-i\delta} \\ 
-s_{12}c_{23}-c_{12}s_{23}s_{13}e^{i\delta}
& c_{12}c_{23}-s_{12}s_{23}s_{13}e^{i\delta}
& s_{23}c_{13} \\ 
s_{12}s_{23}-c_{12}c_{23}s_{13}e^{i\delta}
& -c_{12}s_{23}-s_{12}c_{23}s_{13}e^{i\delta}
& c_{23}c_{13} \\
\end{array}\right)
\left(\begin{array}{c} \nu_1 \\ \nu_2 \\ \nu_3 \end{array} \\ \right)
\; ,
\end{equation}
where $c_{ij} = \cos\theta_{ij}$ and $s_{ij} = \sin\theta_{ij}$. If the
neutrinos are Majorana, there are two extra phases, but these do not
affect oscillations~\cite{phases}.
If $\Delta m^2_{21}$ is responsible for solar neutrino oscillations and
$\Delta m^2_{32}$ for atmospheric neutrino oscillations, then
$|\Delta m^2_{21}| \ll |\Delta m^2_{32}|$. The resulting vacuum oscillation
probabilities for the leading oscillation ($1.267 |\Delta m^2_{21}|
L/E_\nu \ll 1$), appropriate for long-baseline experiments,
are~\cite{bpw80}
\begin{eqnarray}
P(\nu_e \rightarrow \nu_\mu) &=&
\sin^2\theta_{23} \sin^22\theta_{13} \sin^2(1.267 \Delta m^2_{32} L/E_\nu)
\; , \label{Pem}\\
P(\nu_e \rightarrow \nu_\tau) &=&
\cos^2\theta_{23} \sin^22\theta_{13} \sin^2(1.267 \Delta m^2_{32} L/E_\nu)
\; , \label{Pet}\\
P(\nu_\mu \rightarrow \nu_\tau) &=&
\cos^4\theta_{13} \sin^22\theta_{23} \sin^2(1.267 \Delta m^2_{32} L/E_\nu)
\; . \label{Pmt}
\end{eqnarray}
These expressions are analogous to two-flavor vacuum oscillation
probability in Eq.~\ref{vprob}, except that each oscillation channel has
a distinct amplitude that depends on the neutrino mixing parameters.

\subsection{Matter effects}

Electron neutrinos can elastically forward scatter off the electrons in
matter via the charged current interaction~\cite{wolf}. When $\nu_e$
oscillates into either $\nu_\mu$ or $\nu_\tau$, this introduces an 
additional term in the diagonal element of the neutrino 
flavor evolution matrix corresponding to $\nu_e \rightarrow \nu_e$. 
It is useful to define the characteristic matter oscillation 
length $L_0$ as the distance over which the phase of the 
$\nu_e$ wavefunction changes by $2\pi$~\cite{bppw} :
\begin{equation}
L_0 \;=\; \frac{2\pi}{\sqrt{2}G_FN_e} \;\approx\; 
\frac{1.63 \times 10^4}{\rho({\rm g\; cm}^{-3})\;Z/A} \;\;{\rm km}\;,
\label{L0}
\end{equation}
where $N_e = \rho N_0 Z/A$ is the electron density in matter of density
$\rho$, $Z/A$ is the average charge to mass ratio for the electrically
neutral matter, and $N_0$ is Avogadro's number.  Note that, unlike the
vacuum oscillation length $L_V$, the characteristic matter oscillation
length $L_0$ is independent of $E_\nu$. For ordinary rock ($\rho \sim
3$~g~cm$^{-3}$ and $Z/A = 0.5$) $L_0 \approx 10^4$~km. Matter effects
are negligible when $L_0 \gg L$ or $L_0 \gg L_V$. However, matter
effects can be appreciable for trans-Earth experiments, where $L \sim
L_0$, provided $L_0 \lesssim L_V$ which is satisfied if $\Delta
m^2($eV$^2$/c$^4) \lesssim $~E(GeV)/3000.  In practice, for a
trans-Earth experiment, if $E_\mu$ is greater than a few tens of
GeV the muon storage ring becomes too large to have it tilted at a large
angle while preserving relatively long straight sections between the
arcs.  Hence, matter effects will only be important for a trans-Earth
experiment if $\Delta m^2 \leq$~O($10^{-2}$)~eV$^2$/c$^4$. The tilt
angle, distance, and average electron density for some representative
long-baseline experiments from the Fermilab site are given in
Table~\ref{densities_tab}. For distances longer than about 10000~km the
path goes through part of the core (the core diameter is 6960~km and the
Earth's diameter is 12742~km).

The effective $\nu_e$ oscillation length in matter $L_m$ depends on $L_0$ 
and $L_V$. Defining the ratio $R_m \equiv L_m/L_V$, it can be shown 
that~\cite{bppw,kuo} :
\begin{equation}
R_m \;\equiv\; \frac{L_m}{L_V} \;=\; 
\left[ 1 + \left(\frac{L_V}{L_0}\right)^2 - \frac{2L_V}{L_0} \cos 2\theta 
\right]^{-1/2} \; ,
\label{rm}
\end{equation}
and 
\begin{equation}
P(\nu_e \rightarrow \nu_x) \;=\; R_m^2 \; \sin^2 (2\theta) \; 
\sin^2 \left[ \frac{1.267 \; \Delta m^2_{ji} }{R_m}\frac{L}{E_\nu} 
\right] \; .
\label{deneq}
\end{equation}
For antineutrinos, $N_e$ in Eq.~\ref{L0} changes sign; hence matter
effects are different for neutrinos and antineutrinos if
$\cos2\theta \ne 0$. If $\cos2\theta \Delta m^2_{ji} > 0$
(i.e., $\nu_e$ is more closely associated with $\nu_i$ than $\nu_j$),
oscillations of $\nu_e$ are enhanced and oscillations of
$\bar\nu_e$ are suppressed when $L_V \sim L_0$; if $\cos2\theta
\Delta m^2_{ji} < 0$, the situation is reversed. Therefore, matter
effects may distinguish between the two different mass orderings when
$\cos2\theta \ne 0$. 

\begin{figure}
\epsfxsize2.25in
\centerline{\epsffile{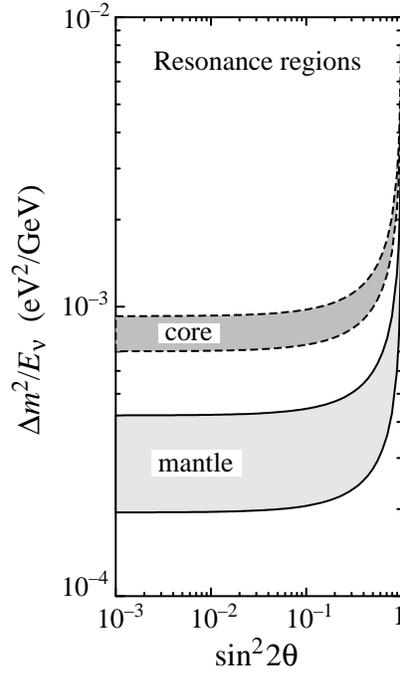}}

\caption{Regions of $\Delta m^2/E_\nu$ versus $\sin^22\theta$ which have
maximal mixing in matter in the Earth. The range of $\Delta m^2/E_\nu$
values is due to the variation of the density in the mantle and the core.
}
\label{res1_fig}
\end{figure}

\begin{figure}
\epsfxsize2.25in
\centerline{\epsffile{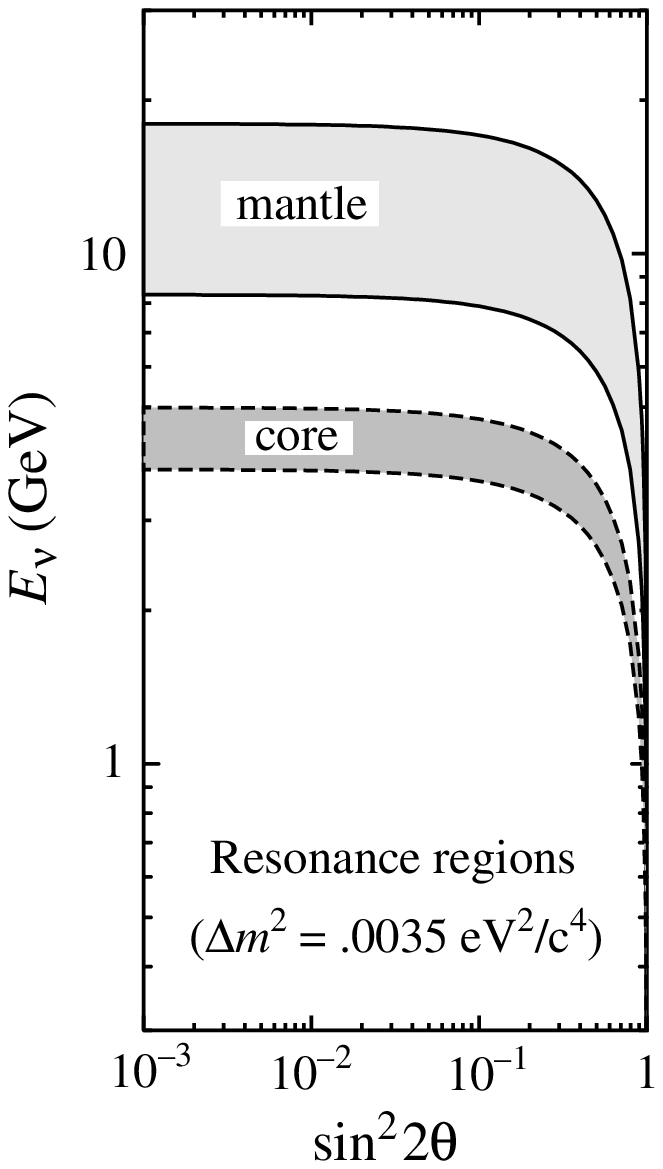}}

\caption{Regions of $E_\nu$ versus $\sin^22\theta$ which have
maximal mixing in matter in the Earth for $\Delta m^2 =
3.5\times10^{-3}$~eV$^2$/c$^4$. The range of $E_\nu$ values
is due to the variation of the density in the mantle and the core.
}
\label{res2_fig}
\end{figure}

If $L_V/L_0 = \cos2\theta$, the oscillations in matter have maximal
mixing~\cite{bppw}. Figure~\ref{res1_fig} shows the values of $\Delta
m^2/E_\nu$ versus $\sin^22\theta$ that give maximal mixing in the mantle
and core of the Earth; similarly, Fig.~\ref{res2_fig} shows values of
$E_\nu$ versus $\sin^22\theta$ that give maximal mixing in the Earth for
$\Delta m^2 = 3.5\times10^{-3}$~eV$^2$/c$^4$, the value favored by the
Super-Kamiokande atmospheric 708-day data~\cite{totsuka}.

\begin{figure}
\epsfxsize4.5in
\centerline{\epsffile{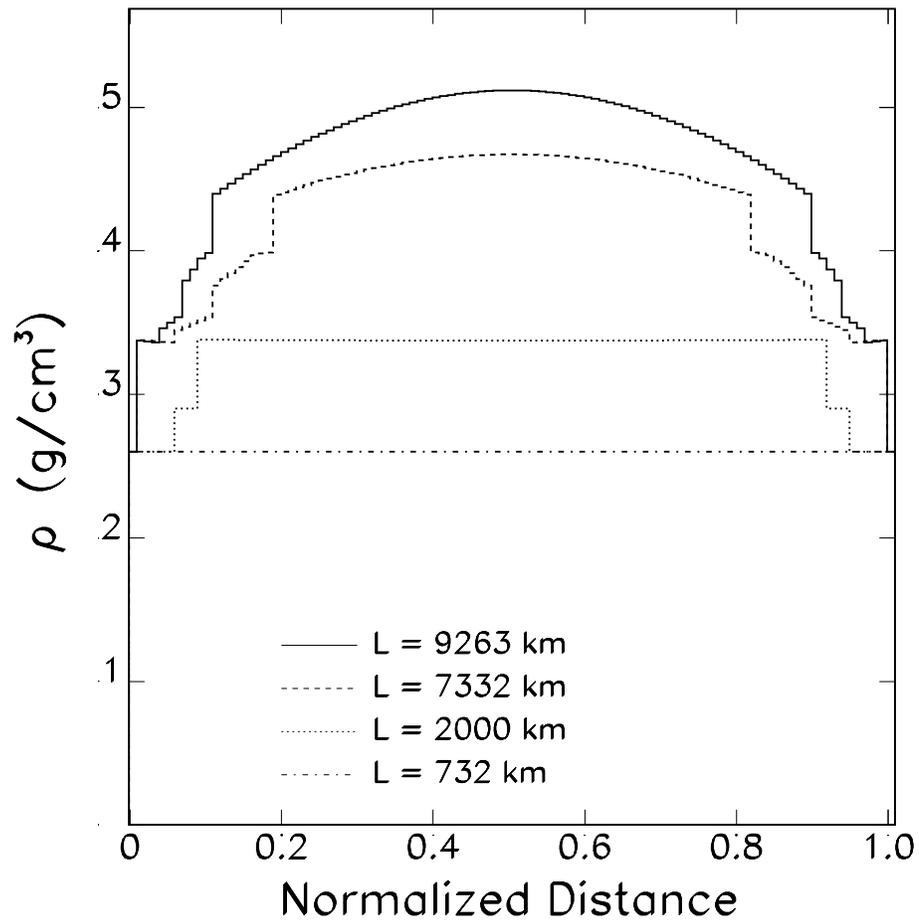}}
\vspace{-1.0cm}
\caption{Density profiles along a selection of chords of length 
$L$ passing through the Earth; the horizontal axis is the fraction of
the total path length.
}
\label{prof}
\end{figure}

\begin{figure}
\epsfxsize4.5in
\centerline{\epsffile{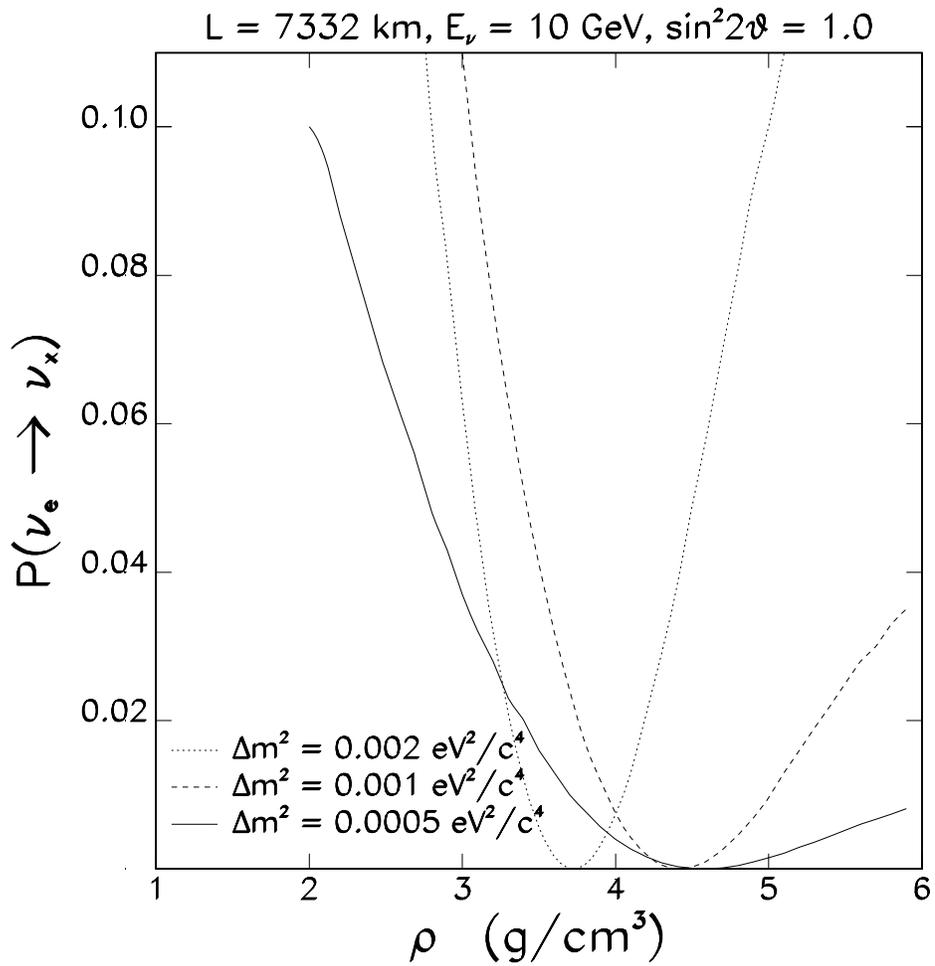}}
\vspace{-1.0cm}
\caption{Electron neutrino disappearance probability 
$P(\nu_e \rightarrow \nu_x$) for $x = \mu$ or $\tau$, shown as a function 
of the assumed matter density for 10~GeV electron neutrinos 
propagating 7332~km through 
the Earth. The curves correspond to the oscillation parameters 
$\sin^2 2\theta = 1$ and $\Delta m^2$ as indicated.
}
\label{prob}
\end{figure}

Note that both the oscillation amplitude and the oscillation length 
depend on $R_m$, and hence on the density $\rho$. Fortunately, outside
of the core the density profile of the Earth is well known, and is
described by the Preliminary Reference Earth 
Model~\cite{earth}. Density profiles along a selection of chords 
passing through the Earth are shown in Fig.~\ref{prof}. 
For $L = 732$~km most of the pathlength is in rock with 
$\rho \sim 2.5$~g~cm$^{-3}$. For $L = 7332$~km most of the pathlength 
is in higher density matter with $\rho > 4$~g~cm$^{-3}$. In some early 
studies~\cite{bppw,parke} matter effects were computed using the
average density  $\rho_{av}$ along the traversed path. This gives a
reasonable estimate in many cases, but can introduce significant
errors in the calculation of $P(\nu_e \rightarrow \nu_x)$ in some
regions of parameter space. To illustrate this, Fig.~\ref{prob}
shows $P(\nu_e \rightarrow \nu_x)$, for $x=\mu$ or $\tau$, versus
$\rho$ calculated using Eq.~\ref{deneq} for 10~GeV neutrinos
propagating 7332~km. The calculation assumes $\sin^2 2\theta = 1$, 
and the results are shown for three different values of $\Delta m^2$. 
For $L = 7332$~km the average density is $\rho_{av} = 4.2$~g/cm$^3$. 
Consider the $\Delta m^2 = 0.001$~eV$^2$/c$^4$ curve, for which 
$L_V = 24800$~km. If $\rho_{av}$ is used we obtain $L_0 = 7810$~km 
and $R_m = 0.30$. The resulting oscillation probability is
$P(\rho_{av}) = 0.002$. However, nearly all of the pathlength is through
matter with densities that correspond to oscillation probabilities
significantly higher than $P(\rho_{av})$. If the true density profile
is used instead of  $\rho_{av}$ we obtain $P = 0.014$. Thus, in this
example, using $\rho_{av}$ leads to a result which underestimates the
oscillation probability by a  factor of 7. Therefore, in the following
sections we do not use $\rho_{av}$ to calculate $P(\nu_e \rightarrow
\nu_x)$, but integrate using the density profile given by the
preliminary Earth model. Extensive analyses have been made of
oscillation effects involving  transmission through the Earth's mantle
and core~\cite{lipari2,lipari,petcov,other_matter,fuller}.

For $\nu_\mu$ oscillations to $\nu_\tau$, there are no matter effects
for simple two-flavor oscillations (although there may be small matter
effects on $\nu_\mu \rightarrow \nu_\tau$ for three
flavors~\cite{pantaleone}). For $\nu_\mu \rightarrow \nu_s$
oscillations, however, muon neutrinos elastically forward scatter off
the quarks and electrons in matter via the neutral current interaction,
whereas sterile neutrinos do not. The matter oscillation length in this
case can be found by replacing $N_e$ in Eq.~\ref{L0} by $-N_n/2$ (by
$N_n/2$ for $\bar\nu_\mu \rightarrow \bar\nu_s$
oscillations)~\cite{bdppw}; $Z$ in Eq.~\ref{L0} is then replaced by
$Z-A$ ($A-Z$). If the mixing of atmospheric $\nu_\mu$ is nearly maximal
(as suggested by atmospheric neutrino data), then Eqs.~\ref{rm} and
\ref{deneq} imply that the amplitude of $\nu_\mu \rightarrow \nu_s$ and
$\bar\nu_\mu \rightarrow \bar\nu_s$ oscillations will both be suppressed
by matter effects if $L_V \agt L_0$, and by the same amount since
$\cos2\theta \simeq 0$.

\section{$\nu_\mu \rightarrow \nu_\tau$ and $\nu_\mu \rightarrow \nu_{\lowercase{s}}$ Oscillations}

The currently favored explanation for the Super-Kamiokande atmospheric
neutrino results is that muon neutrinos are oscillating primarily into
either tau neutrinos or sterile neutrinos, with the oscillation
parameters given by $\sin^2 2\theta \sim 1$ and $\Delta m^2$ in the
approximate range 0.002~eV$^2$/c$^4$ to
0.006~eV$^2$/c$^4$~\cite{superk,totsuka}. Searches for neutrino oscillations at accelerators are based on either the disappearance of the initial neutrino flavor, or the appearance of a neutrino flavor not originally within the neutrino beam. If the current interpretation of the Super-Kamiokande data is correct, the next generation of approved long baseline accelerator experiments~\cite{K2K,MINOS} should confirm the existence of neutrino oscillations and make the first laboratory measurements of the oscillation parameters. This will happen in the period before a first muon storage ring neutrino source could be built. It is likely therefore that the main atmospheric neutrino oscillation physics goals in the muon storage ring neutrino source era would be to make very precise measurements of the oscillation parameters, and determine whether there is a small $\nu_\mu \rightarrow \nu_s$ ($\nu_\mu \rightarrow \nu_\tau$) component in a dominantly
$\nu_\mu \rightarrow \nu_\tau$ ($\nu_\mu \rightarrow \nu_s$)
signal. CC measurements can distinguish
between $\nu_\mu \rightarrow \nu_\tau$ and $\nu_\mu \rightarrow \nu_s$
oscillations in two ways: (i) direct appearance of taus for $\nu_\mu
\rightarrow \nu_\tau$~\cite{hall} and (ii) a different $\nu_\mu
\rightarrow \nu_s$ transition probability due to matter effects when
compared to $\nu_\mu \rightarrow \nu_\tau$. NC/CC measurements of
$\pi^0$ production can also distinguish $\nu_\mu \rightarrow \nu_s$
from $\nu_\mu \rightarrow \nu_\tau$~\cite{smirnov}.

\subsection{Fermilab $\rightarrow$ Gran Sasso}

\begin{figure}
\epsfxsize4.5in
\centerline{\epsffile{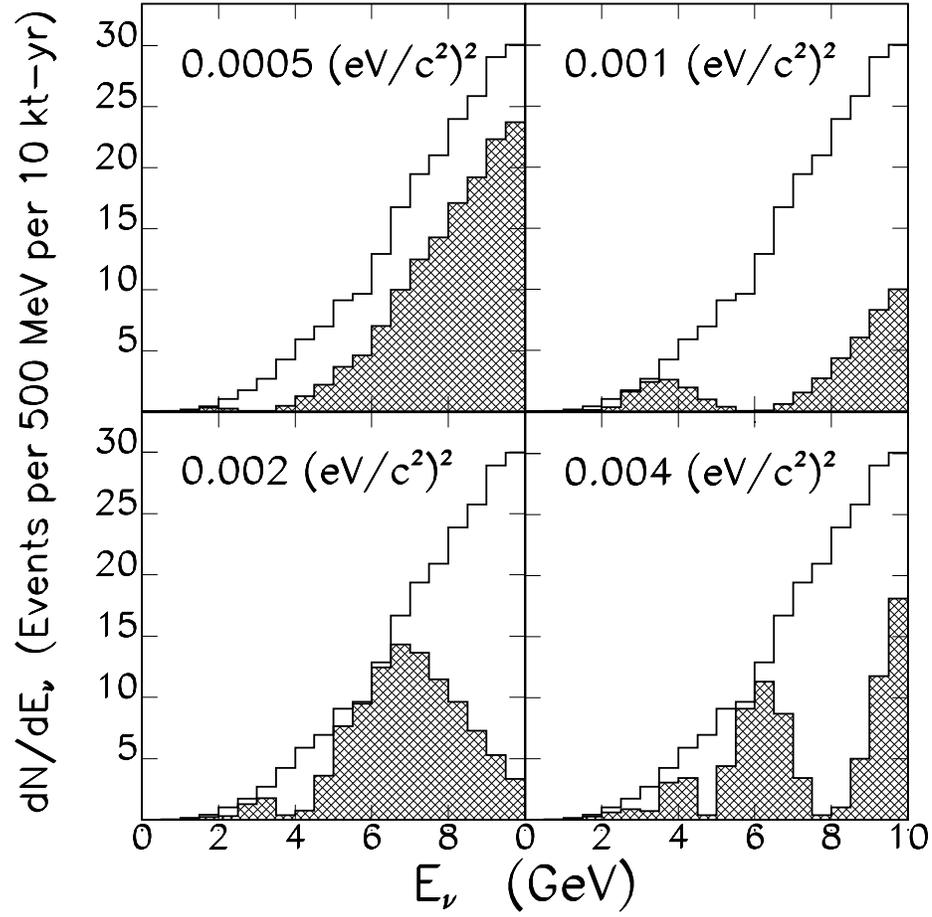}}
\vspace{-1.0cm}
\caption{Modified $\nu_\mu$ CC interaction spectra for 
a 10~GeV muon storage ring neutrino source located at FNAL and a 
detector at the Gran Sasso underground laboratory, shown 
for several values of the oscillation parameter $\Delta m^2$, 
assuming $\sin^2 2\theta = 1$. In each of the four panels 
the upper curves show the unmodulated spectrum, and the 
lower curves the modulated spectrum corresponding to the 
indicated $\Delta m^2$.
}
\label{numu_gs}
\end{figure}

Consider first a 10~GeV muon storage ring at Fermilab with the neutrino beam-forming straight section pointing at Gran Sasso ($L = 7332$~km). The predicted CC event rates are listed as a function of $E_\mu$ and $\Delta m^2$ in Table~\ref{osc_rate_tab}. The predicted $\nu_\tau$ appearance event rates arising from $\nu_\mu \rightarrow \nu_\tau$ oscillations with $\sin^2 2\theta = 1$ and $\Delta m^2$ in the range from 0.002~eV$^2$/c$^4$ to 0.006~eV$^2$/c$^4$ are only of order 1~event per kt-yr. This event rate is too low to enable a precise $\nu_\tau$ appearance measurement in, for example, a 1~kt hybrid emulsion detector. However, an oscillation measurement could be made by a $\nu_\mu$ disappearance experiment. In the absence of oscillations about 220~$\nu_\mu$ CC events would be
expected per year in a 10~kt detector, with an event energy distribution
peaking at $\sim 10$~GeV. In the presence of oscillations the predicted $\nu_\mu$ CC event rate (Table~\ref{osc_rate_tab}) and the corresponding event energy distribution (Fig.~\ref{numu_gs}) are both sensitive
to $\Delta m^2$ (provided $\eta \sim 1$). However in general, for a given $\Delta m^2$, the average oscillation probability will depend on whether $\nu_\mu \rightarrow \nu_\tau$ or $\nu_\mu \rightarrow \nu_s$ or a mixture of the two oscillations is being observed. Hence, to avoid a large uncertainty in the extracted value of $\Delta m^2$ due to an uncertainty in which oscillation mode is being observed, it is desirable that a trans-Earth baseline experiment is sensitive to both $\nu_\mu$ disappearance and $\nu_\tau$ appearance. We conclude that the muon storage ring beam energy needs to be higher than 10~GeV to facilitate a $\nu_\tau$ appearance measurement.
Consider next a 20~GeV muon storage ring at Fermilab with the neutrino
beam-forming straight section pointing at Gran Sasso ($L = 7332$~km).
The $\nu_\tau$ CC rate corresponding to the favored region of $\nu_\mu \rightarrow \nu_\tau$ parameter space is now sufficiently large to facilitate a $\nu_\tau$ appearance measurement. In the absence of oscillations about 1900~$\nu_\mu$ CC events would be
expected per year in a 10~kt detector. In the presence of $\nu_\mu \rightarrow \nu_\tau$ oscillations with $\sin^2 2\theta = 1$
and $\Delta m^2 =$~0.002~eV$^2$/c$^4$ (0.006~eV$^2$/c$^4$) only 370 (1300) $\nu_\mu$ CC events per 10~kt-yr would be expected, together with 43 (14) $\nu_\tau$ CC events per year in a 1~kt detector.
The corresponding event rates for $\nu_\mu \rightarrow \nu_s$ oscillations are 1400 (1200) $\nu_\mu$ CC events per 10~kt-yr and no $\nu_\tau$ CC events. With these trans-Earth event rates at a 20~GeV storage ring a $\nu_\mu$ disappearance measurement would enable the oscillation probability to be measured with a precision of a few percent and, for a known mixture of $\nu_\mu \rightarrow \nu_\tau$ and $\nu_\mu \rightarrow \nu_s$ oscillations, would enable $\Delta m^2$ to be determined with a statistical precision of a few percent. In addition, a $\nu_\tau$ appearance measurement would enable the $\nu_\mu \rightarrow \nu_\tau$ oscillation probability to be measured with a statistical precision which depends on $\Delta m^2$, but is typically about 20\%. Finally, a comparison of the appearance and disappearance results would enable a $\nu_\mu \rightarrow \nu_s$ contribution at the few times 10\% level to be observed in a predominantly $\nu_\mu \rightarrow \nu_\tau$ signal, or an approximately 10\% $\nu_\mu \rightarrow \nu_\tau$ contribution to be measured in a predominantly $\nu_\mu \rightarrow \nu_s$ signal.

\subsection{Fermilab $\rightarrow$ Soudan}

\begin{figure}
\epsfxsize4.5in
\centerline{\epsffile{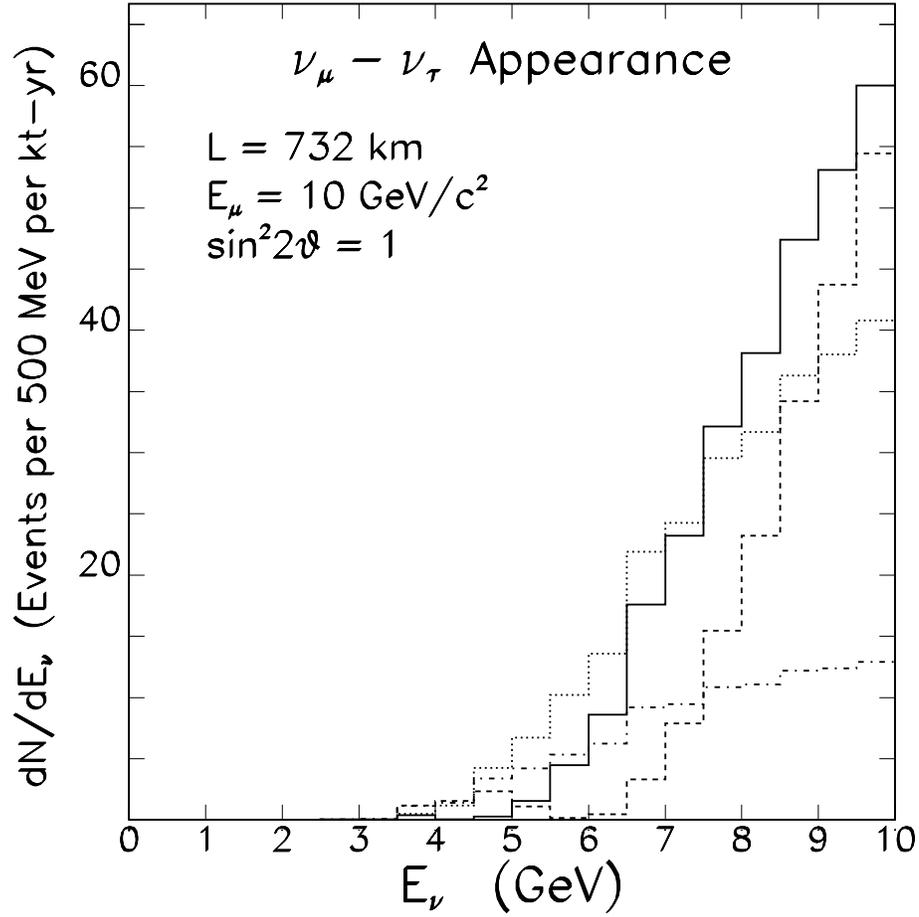}}
\vspace{-1.0cm}
\caption{Predicted $\nu_\tau$ appearance CC interaction spectra for 
a 10~GeV muon storage ring neutrino source located at FNAL and a
detector at the Soudan mine, shown for several values of the oscillation parameter $\Delta m^2$, assuming $\sin^2 2\theta = 1$.  The curves correspond to $\Delta m^2 = 0.02$~eV$^2$/c$^4$ (dotted), 0.015~eV$^2$/c$^4$ (solid), 0.01~eV$^2$/c$^4$ (broken), and 0.005~eV$^2$/c$^4$ (dot-dashed).
}
\label{numu_tau}
\end{figure}

Consider a 10~GeV muon storage ring at Fermilab with the neutrino
beam-forming straight section pointing at Soudan ($L = 732$~km).
Predicted event rates are listed in Table~\ref{osc_rate_tab}. In the absence of oscillations about 22200~$\nu_\mu$ CC events would be
expected per year in a 10~kt detector. In the presence of either $\nu_\mu \rightarrow \nu_\tau$ or $\nu_\mu \rightarrow \nu_s$ oscillations with $\sin^2 2\theta = 1$
and $\Delta m^2 =$~0.002~eV$^2$/c$^4$ (0.006~eV$^2$/c$^4$)
only 20500 (11700) $\nu_\mu$ CC events per 10~kt-yr would be expected. In addition, for $\nu_\mu \rightarrow \nu_\tau$ oscillations 20 (150) $\nu_\tau$ CC events per year would be expected in a 1~kt detector.
Note that the shape of the
$\nu_\tau$ CC interaction energy distribution (Fig.~\ref{numu_tau}) exhibits some dependence on $\Delta m^2$, although a detector with good energy resolution would be needed to
exploit this dependence.
We conclude that a storage ring with a beam energy as low as 10~GeV might be of interest for a Fermilab $\rightarrow$ Soudan experiment, particularly if $\Delta m^2$ is at the upper end of the currently favored region. In this case the average oscillation probability could be measured with a statistical precision of better than 1\% from the disappearance measurement, and the resulting $\Delta m^2$ be determined with a precision of $\sim$1\%. Note that, since matter effects are small, there is no additional uncertainty on $\Delta m^2$ arising from an imprecise knowledge of the oscillation mode ($\nu_\mu \rightarrow \nu_\tau$ or $\nu_\mu \rightarrow \nu_s$). Finally, with $\Delta m^2 =$~0.006~eV$^2$/c$^4$, a comparison of the $\nu_\mu$ disappearance and $\nu_\tau$ appearance measurements would enable a few percent contribution from $\nu_\mu \rightarrow \nu_\tau$ oscillations to be measured in a predominantly $\nu_\mu \rightarrow \nu_s$ signal, or a few times 10\% $\nu_\mu \rightarrow \nu_s$ contribution to be measured in a predominantly $\nu_\mu \rightarrow \nu_\tau$ signal.

\begin{figure}
\epsfxsize4.5in
\centerline{\epsffile{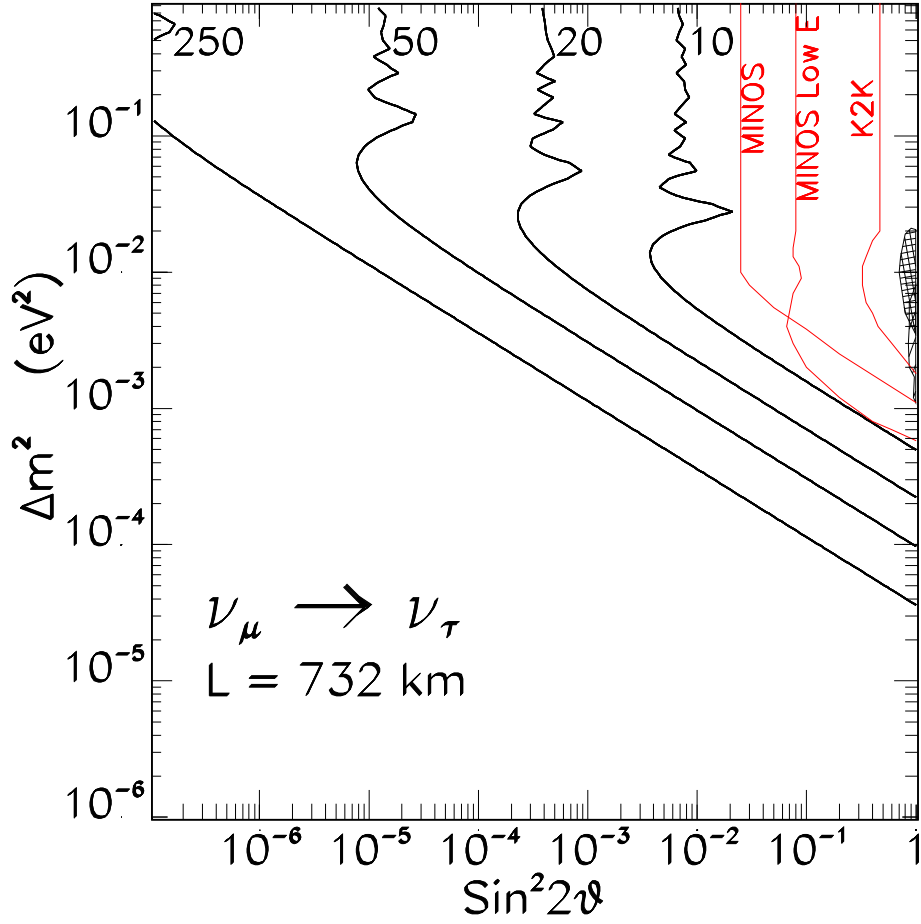}}
\vspace{-1.0cm}
\caption{Single event per kt-yr contours calculated for 
$\nu_\mu \rightarrow \nu_\tau$ appearance 732~km downstream of a muon
storage ring neutrino source. From left to right, the contours
correspond to unpolarized muons with energies $E_\mu =$~250, 50, 20, and
10~GeV. The shaded areas correspond
to the Kamiokande and Super-Kamiokande allowed region of parameter
space. Also shown are the expected regions of sensitivity for the MINOS
and K2K experiments (as labelled).
}
\label{numu_t2}
\end{figure}

\begin{figure}
\epsfxsize4.5in
\centerline{\epsffile{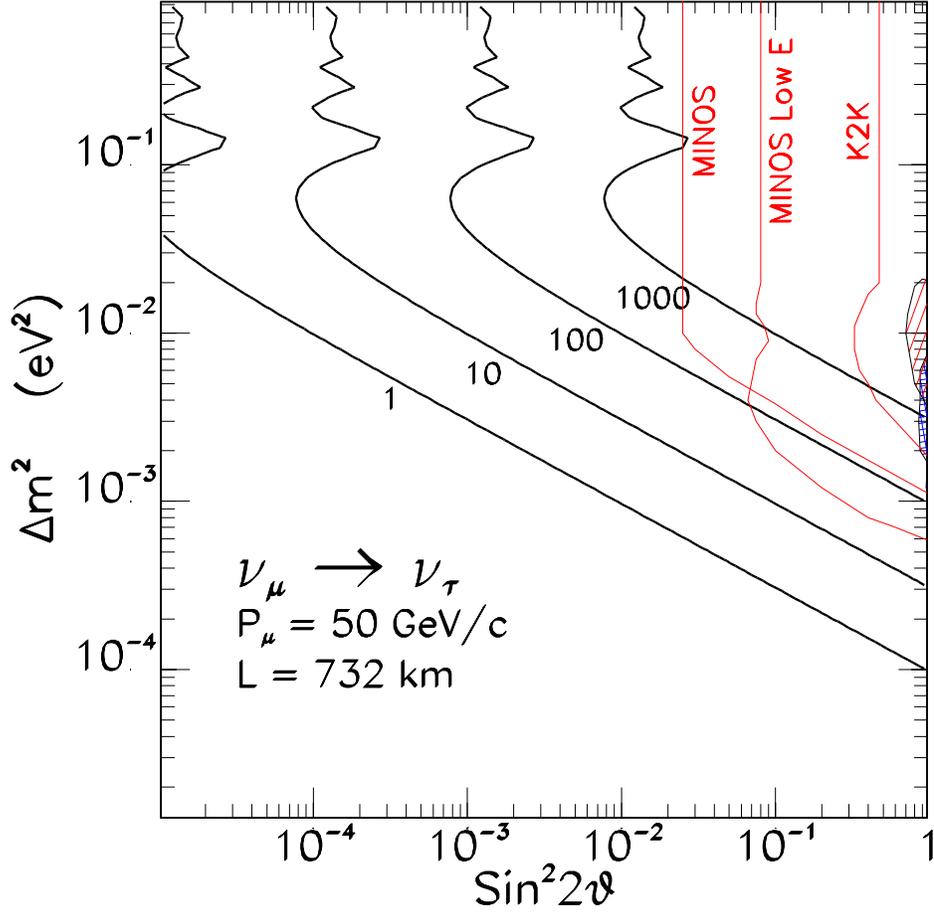}}
\vspace{-1.0cm}
\caption{Contours of constant $\nu_\tau$ CC interaction event rate
for $\nu_\mu \rightarrow \nu_\tau$ appearance 732~km downstream of
a 50~GeV muon storage ring neutrino source. From left to right, the
contours correspond to event rates of 1, 10, 100, and 1000 per kt-yr.
The shaded areas correspond to the
Kamiokande and Super-Kamiokande allowed region of parameter space. Also
shown are the expected regions of sensitivity for the MINOS and K2K
experiments (as labelled).
}
\label{numu_t1}
\end{figure}

In Fig.~\ref{numu_t2} single $\nu_\tau$ CC event per kt-yr contours in the ($\sin^2 2\theta$, $\Delta m^2$)-plane are shown for $L = 732$~km as a function of $E_\mu$. The sensitivity of the $\nu_\tau$ appearance measurement increases with increasing muon storage ring beam energy. Consider the sensitivity for a 50~GeV muon storage ring. The dependence of the $\nu_\tau$ CC rates on $\sin^2 2\theta$ and
$\Delta m^2$ is shown in Fig.~\ref{numu_t1}.
The predicted event rates range from several hundred events per kt-yr
at the lower end of the Super-Kamiokande allowed region of parameter space, to several thousand events per kt-yr at the upper end of the favored region (Table~\ref{osc_rate_tab}). Hence the $\nu_\mu \rightarrow \nu_\tau$ oscillation probability could be measured with a precision of a few percent, corresponding to a determination of $\Delta m^2$ with a precision of a few percent. However, the predicted oscillation probabilities are small, and the statistical precision of the disappearance
measurement is less than the corresponding precision for the
$E_\mu = 10$~GeV case. Note that about 2~million~$\nu_\mu$ CC events would be expected per year in a 10~kt detector. The sensitivity of the disappearance measurement may therefore be limited by systematics. This needs further study.
We conclude that, although the statistical sensistivity of the $\nu_\tau$ appearance measurement improves with increasing $E_\mu$, the optimum choice of muon beam energy depends upon whether $\Delta m^2$ is in the lower or upper part of the favored region of parameter space, and on at what level the higher energy experiments would be limited by systematics.

\subsection{$L = 7332$~km versus 732~km}

The relative performance of experiments at long ($L = 732$~km) and very long ($L = 7332$~km) baselines depends on both $\Delta m^2$ and the oscillation mode ($\nu_\mu \rightarrow \nu_\tau$or $\nu_\mu \rightarrow \nu_s$). Based on the results summarized in Table~\ref{osc_rate_tab} we find that:
\begin{description}
\item{Case 1:} $0.002 < \Delta m^2 < 0.004$~eV$^2$/c$^4$ and $\nu_\mu \rightarrow \nu_\tau$ oscillations dominate. An $L = 732$~km experiment with $E_\mu = 10$-20~GeV, or an $L = 7332$~km experiment with $E_\mu = 20$~GeV both seem interesting.
A higher energy $L = 732$~km experiment (e.g.\ $E_\mu = 50$~GeV) would provide a higher statistics measurement of the $\nu_\mu \rightarrow \nu_\tau$ oscillation probability, but the oscillation probability is small and the measurement precision may be dominated by systematics.
\item{Case 2:} $0.004 < \Delta m^2 < 0.006$~eV$^2$/c$^4$ and $\nu_\mu \rightarrow \nu_\tau$ oscillations dominate.
An $L = 732$~km experiment with $E_\mu = 10$--20~GeV would provide a good $\nu_\mu$ disapearance measurement and a good $\nu_\tau$ appearance measurement. With a larger $E_\mu$ (e.g.~50~GeV) the unoscillated neutrino CC rate becomes large and the oscillation measurements may be limited by systematics. The trans-Earth baseline option with $E_\mu = 20$~GeV yields less statistical precision than the $L = 732$~km option with $E_\mu = 20$~GeV.
\item{Case 3:} $\nu_\mu \rightarrow \nu_s$ oscillations dominate. In this case a very sensitive $\nu_\mu$ disapearance measurement is important. A low energy (10~GeV) storage ring pointing at Soudan seems attractive, particularly if $\Delta m^2$ is in the upper half of the favored region. An $L = 7332$~km baseline experiment seems to be less attractive than an $L = 732$~km baseline experiment for large $\Delta m^2$, but is  complementary to the shorter baseline experiment for small $\Delta m^2$, and would enable matter effects to be measured.
\end{description}

\section{$\nu_{\lowercase{e}} \rightarrow \nu_\mu$ Oscillations}

If confirmed, the LSND neutrino oscillation results necessarily imply
the existence of $\nu_e \rightarrow \nu_\mu$ oscillations, with $0.3 <
\Delta m^2_{LSND} < 2.0$~eV$^2$/c$^4$. Oscillation limits from the
Karmen experiment~\cite{KARMEN} are no longer in conflict with the LSND
effect.  In addition, the solar neutrino results can also be interpreted
in terms of $\nu_e \rightarrow \nu_\mu$ and/or $\nu_e \rightarrow
\nu_\tau$ oscillations either within the framework of the MSW effect
with $\Delta m^2_{sun} \sim 5\times 10^{-5}$~eV$^2$/c$^4$ and $\sin^2
2\theta \sim 0.8$ (large angle solution) or $\Delta m^2_{sun} \sim
5\times 10^{-6}$~eV$^2$/c$^4$ and $\sin^2 2\theta \sim 5\times10^{-3}$
(small angle solution)~\cite{bks1,bks2,newvalle}, or within the
framework of vacuum oscillations~\cite{justso} with $\Delta m^2_{sun}
\sim 0.65-8.6\times10^{-10}$~eV$^2$/c$^4$~\cite{global,bks1}. In the
context of three neutrinos, although the primary oscillation of
atmospheric neutrinos is $\nu_\mu \rightarrow \nu_\tau$, a small
$\nu_\mu \rightarrow \nu_e$ component is not excluded.  For $\Delta
m^2_{atm} \sim 3.5\times10^{-3}$~eV$^2$/c$^4$, a measurement of $\nu_e
\rightarrow \nu_\mu$ will place strong limits on $\sin^2\theta_{23}
\sin^22\theta_{13}$ (see Eq.~\ref{Pem})~\cite{three,global}.

The next generation of approved neutrino experiments include
MiniBooNE~\cite{boone} at Fermilab, and KamLAND~\cite{kam} at the
Kamioka Mine, Japan. MiniBooNE, which will consist of a 445~ton fiducial
volume liquid scintillator detector 500~m downstream of a $\nu_\mu$
source, with $E_\nu \sim 0.75$~GeV, should confirm or refute the
neutrino oscillation interpretation of the LSND results.  KamLAND, which
will consist of a 1~kt liquid scintillator detector 140~km and 200~km
from reactors in Japan, should observe a $\nu_e$ disappearance signal if
the solar neutrino results are due to oscillations corresponding to the
large angle MSW solution.  Both MiniBooNE and KamLAND nominally begin
operation in $\sim 2001$.  In the following we consider the $\nu_e
\rightarrow \nu_\mu$ physics potential of a muon storage ring neutrino
source in the post-MiniBooNE/KamLAND era.

\subsection{Fermilab $\rightarrow$ Soudan}

\begin{figure}
\epsfxsize4.5in
\centerline{\epsffile{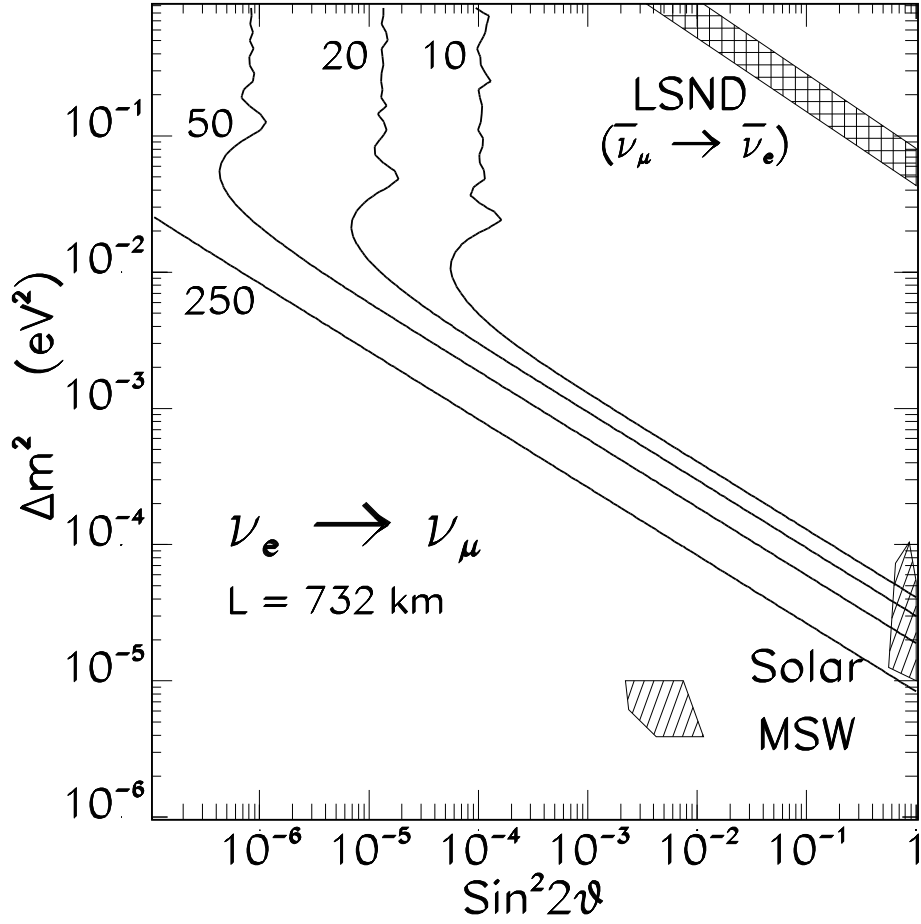}}
\vspace{-1.0cm}
\caption{Single event per 10~kt-yr contours calculated for 
$\nu_e \rightarrow \nu_\mu$ appearance 732~km downstream of 
a muon storage ring neutrino source. From left to right, the 
contours correspond to unpolarized muons with energies 
$E_\mu =$~250, 50, 20, and 10~GeV. The regions favored by the LSND
results and the MSW solar neutrino oscillation solutions are also shown.
}
\label{nue_mu1}
\end{figure}

\begin{figure}
\epsfxsize4.5in
\centerline{\epsffile{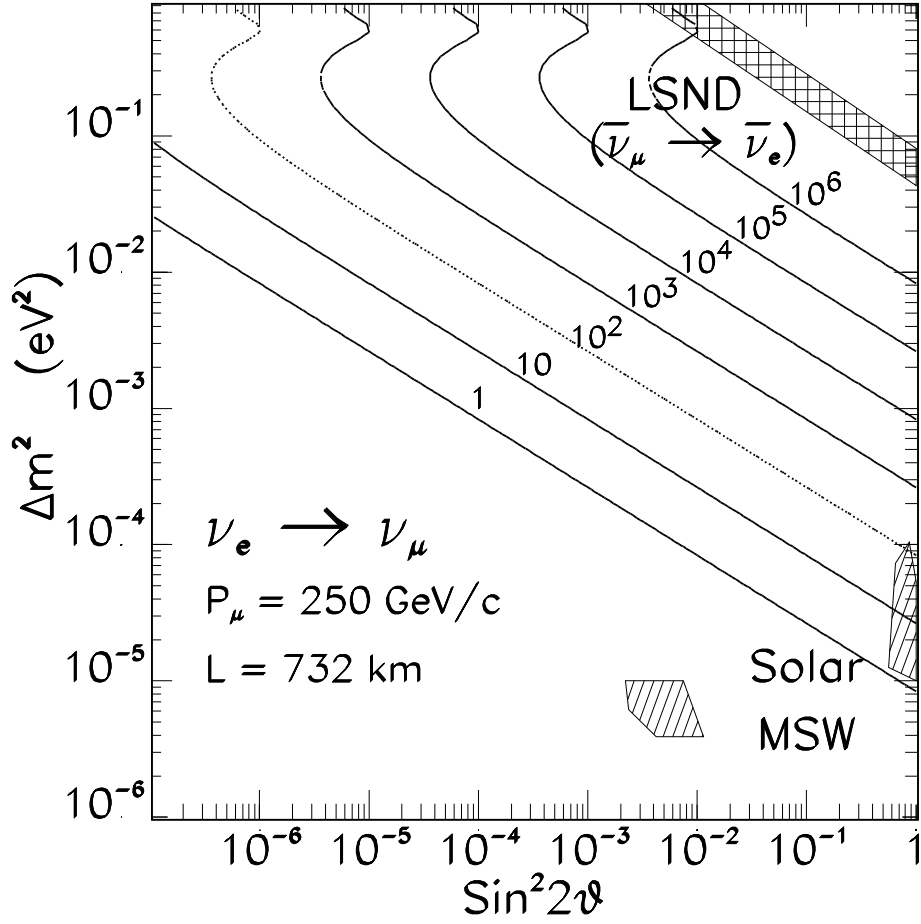}}
\vspace{-1.0cm}
\caption{Contours of constant $\nu_\mu$ CC interaction event rate 
for $\nu_e \rightarrow \nu_\mu$ appearance 732~km downstream of a
250~GeV muon storage ring neutrino source. From left to right, the
contours correspond to event rates of 1, 10, $10^2$, $10^3$, $10^4$,
$10^5$ and $10^6$ per 10~kt-yr. The regions favored by the LSND
results and the MSW solar neutrino oscillation solutions are also shown.
}
\label{nue_mu2}
\end{figure}

Figure~\ref{nue_mu1} shows single event per year contours in 
the ($\sin^2 2 \theta, \Delta m^2$)-plane for $\nu_\mu$ CC 
interactions in a 10~kt detector 732~km downstream of the muon 
storage ring neutrino source described in Section~II.A, where 
the $\nu_\mu$ arise from $\nu_e \rightarrow \nu_\mu$ oscillations. 
The contours are shown for several values of $E_\mu$ from 10~GeV up to 250~GeV.
The upper part of the MSW large mixing angle solution is within the
single-event per year boundary for $E_\mu \ge 10$~GeV. The sensitivity
improves with increasing $E_\mu$. The event rate is shown in the
$\sin^2 2 \theta$--$\Delta m^2$ plane in Fig.~\ref{nue_mu2} for 
$E_\mu = 250$~GeV. To completely cover the parameter space corresponding 
to the MSW large mixing angle solution would require larger muon beam 
currents than those of the scenario described in Section~II.A. It should 
be noted that the full muon collider front-end described in 
Ref.~\cite{status_report}, would produce a factor of 3.3 more muons. 
The resulting neutrino flux would just about enable the entire 
MSW large mixing angle solution to be covered by a 50~GeV storage ring 
and a few years of running with, for example, a 20~kt detector having a 
detection efficiency of 50\%, provided background rates are negligible. 
The small angle MSW solution is out of reach of a muon storage ring
neutrino source, as currently envisioned. Hence if KamLAND rules out the
large angle solution, then solar neutrino oscillations would provide 
little motivation for a Fermilab $\rightarrow$ Soudan 
$\nu_e \rightarrow \nu_\mu$ appearance experiment. On the other hand, 
should KamLAND obtain a positive signal for solar neutrino oscillations 
corresponding to the large angle MSW solution, a Fermilab $\rightarrow$ 
Soudan muon storage ring neutrino source experiment to search for 
$\nu_e \rightarrow \nu_\mu$ appearance might be feasible, but would 
clearly be challenging and would probably require the full muon 
collider muon source for several years. If this becomes the only way 
to discover whether the solar neutrino oscillations are due to 
$\nu_e \rightarrow \nu_\mu$ or due to $\nu_e \rightarrow \nu_\tau$ or 
$\nu_s$, then this experiment would certainly be worthy of serious 
consideration.

Independent of the KamLAND results, a positive MiniBooNE result would
motivate precise measurements of $\nu_e \rightarrow \nu_\mu$
oscillations to test whether or not $P(\nu_e \rightarrow \nu_\mu)$ =
$P(\nu_\mu \rightarrow \nu_e)$ ($T$ invariance), test the oscillation
phenomenon with a very different $L$ and $E_\nu$, search for a $\nu_e
\rightarrow \nu_s$ contribution, etc. A Fermilab $\rightarrow$ Soudan
$\nu_e \rightarrow \nu_\mu$ experiment would enable many thousands of
$\nu_\mu$ appearance CC interactions (due to LSND-scale oscillations)
to be measured (see Figs.~\ref{nue_mu1} and \ref{nue_mu2}). Such
measurements could also test the existence of $\nu_e \rightarrow
\nu_\mu$ oscillations at the atmospheric $\Delta m^2$ scale ($\sim
3.5\times10^{-3}$~eV$^2$/c$^4$), which allows a precise determination of
the three-neutrino mixing parameter
$\cos^2\theta_{23}\sin^22\theta_{13}$.

\subsection{Matter effects}

\begin{figure}
\vspace*{.5in}
\epsfxsize4.5in
\centerline{\epsffile{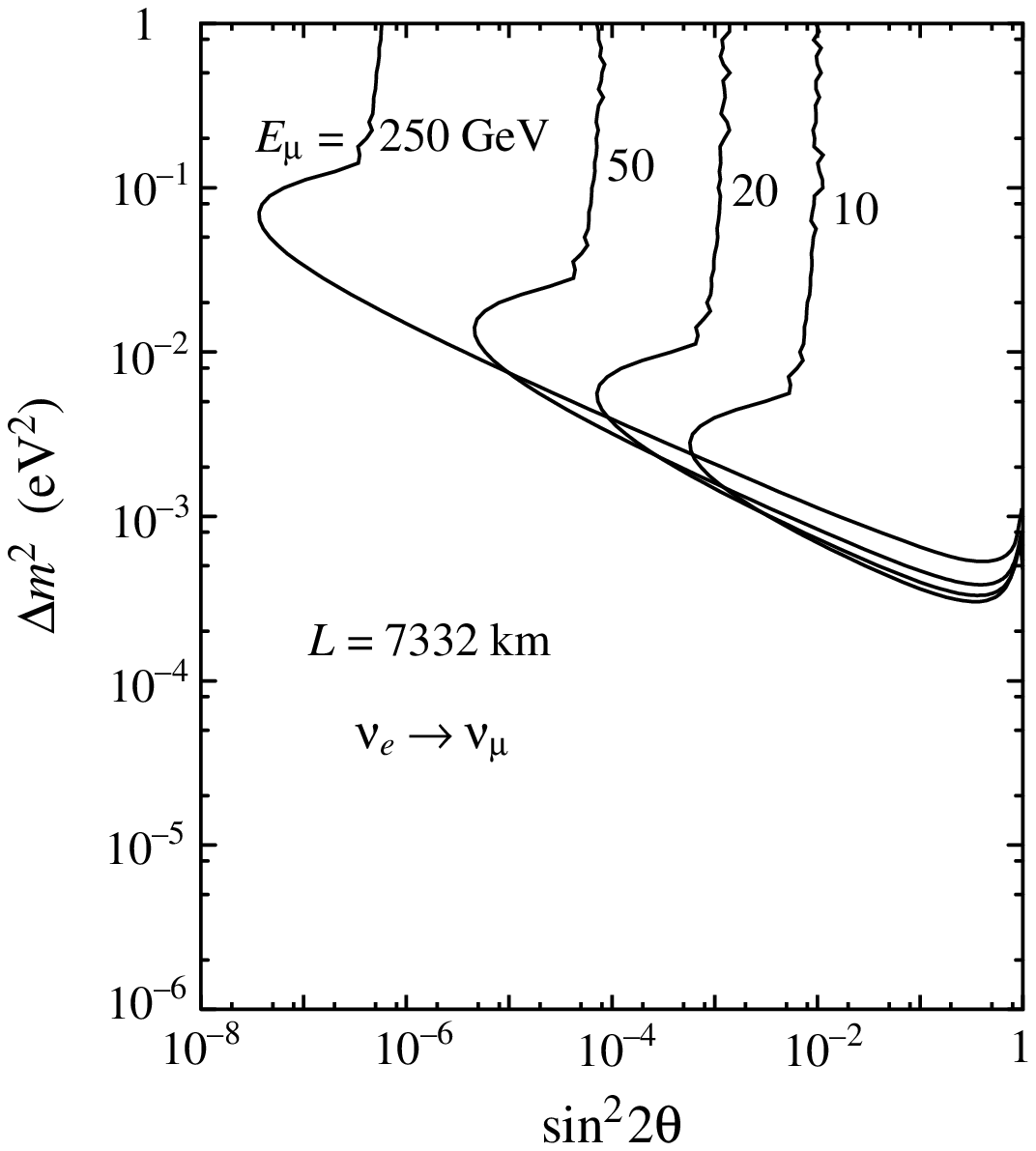}}
\vspace{1cm}
\caption{Single event per 10~kt-yr contours calculated for 
$\nu_e \rightarrow \nu_\mu$ appearance 7332~km downstream of a muon
storage ring neutrino source, assuming $\Delta m^2 >0$. From left to
right, the contours correspond to unpolarized muons with energies 
$E_\mu =$~250, 50, 20, and 10~GeV.
}
\label{nue_mu4}
\end{figure}

\begin{figure}
\vspace*{.5in}
\epsfxsize4.5in
\centerline{\epsffile{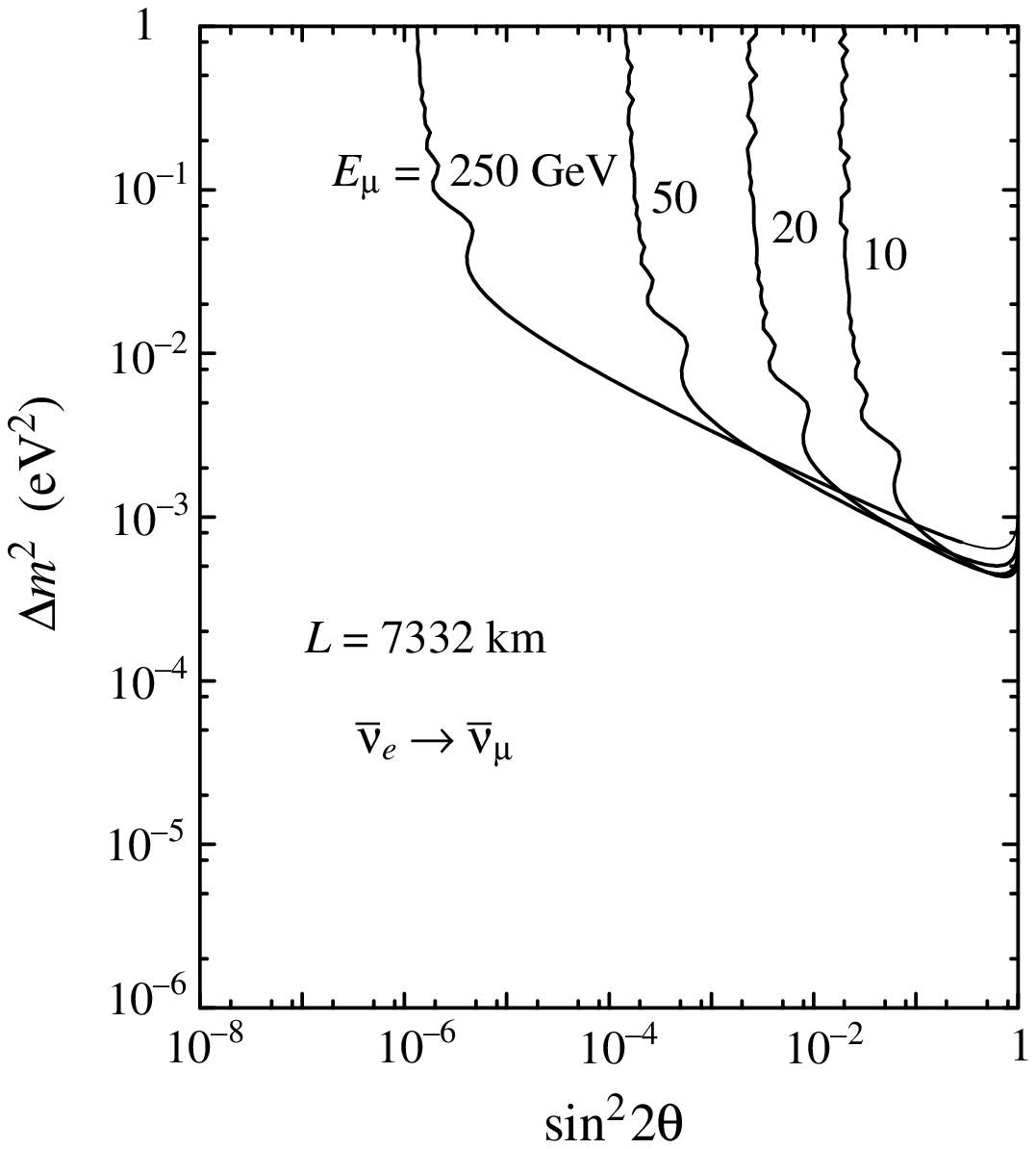}}
\vspace{1cm}

\caption{(a) Single event per 10~kt-yr contours calculated for 
$\bar\nu_e \rightarrow \bar\nu_\mu$ appearance 7332~km downstream of a
muon storage ring neutrino source, assuming $\Delta m^2 >0$. From left
to right, the contours correspond to unpolarized muons with energies 
$E_\mu =$~250, 50, 20, and 10~GeV.
}
\label{nue_mu5}
\end{figure}

Figure~\ref{nue_mu4} shows single event per year contours in the
($\sin^2 2 \theta, \Delta m^2$)-plane for $\nu_\mu$ CC interactions in
a 10~kt detector 7332~km downstream of the muon storage ring neutrino
source described in Section~II.A, where the $\nu_\mu$ arise from $\nu_e
\rightarrow \nu_\mu$ oscillations with $\Delta m^2 >0$. The contours are
shown for $E_\mu=$~10, 20, 50, and 250~GeV. Figure~\ref{nue_mu5} shows
the corresponding contours for $\bar\nu_e \rightarrow \bar\nu_\mu$
oscillations. Compared with the equivalent contours for $L = 732$~km
(Fig.~\ref{nue_mu1}) the lower event statistics suppress the sensitivity
at large $\Delta m^2$ and small $\sin^2 2\theta$, and matter effects
(Eq.~\ref{deneq}) suppress the sensitivity at small $\Delta m^2$ and
large $\sin^2 2\theta$. It is interesting to consider whether the matter
effects might be measured by a trans-Earth muon storage ring neutrino
source $\nu_e \rightarrow \nu_\mu$ experiment searching for wrong-sign
muons.

Note that if MiniBooNE confirms the LSND evidence for oscillations, 
then $\Delta m^2$ will be too large to result in a significant 
modification of the oscillation probability due to matter effects. 
However, if KamLAND obtains a positive $\nu_e$ disappearance result 
corresponding to the large mixing angle MSW solution for 
solar neutrino oscillations, then the $\Delta m^2$ will be in the 
right range to produce a significant modification of the oscillation 
probability due to matter. As an example, consider a 
storage ring with $E_\mu = 20$~GeV, and assume that $\sin^2 2\theta = 1$. 
Then if $\Delta m^2 = 1 \times 10^{-4}$~eV$^2$/c$^4$, in the absence of 
matter effects $\sim 11$ wrong sign muons (from $\nu_e \rightarrow \nu_\mu$ 
oscillations) would be expectected per 10~kt-yr. With matter effects, 
the number of wrong sign muons is reduced to $\sim 0.6$ per 10~kt-yr. 
Hence, an experiment to measure matter effects, by comparing the 
wrong sign muon rates in a long baseline (e.g $L = 732$~km) experiment
with the  corresponding rates in a trans-Earth (e.g $L = 7332$~km)
experiment might be feasible if the solar neutrino results are due to
$\nu_e \rightarrow \nu_\mu$ oscillations with $\sin^2 2\theta \sim 1$ 
and $\Delta m^2$ at the upper end of the region of allowed parameter 
space for the MSW large mixing angle solution. This experiment would 
clearly be challenging, and would probably need a more intense neutrino 
source than described in Section~II.A (for example, the full muon collider 
muon source utilized for several years with a 20~kt detector).

If there is a small $\nu_\mu \rightarrow \nu_e$ and $\bar\nu_\mu
\rightarrow \bar\nu_e$ component to atmospheric neutrino oscillations,
then $\nu_e \rightarrow \nu_\mu$ and $\bar\nu_e \rightarrow \bar\nu_\mu$
oscillations are also expected; a trans-Earth experiment could test for
different matter effects acting on these two oscillation
channels. Table~\ref{emu_tab} shows, for a
Fermilab $\rightarrow$ Gran Sasso experiment ($L=7332$~km), 
the $\nu_e \rightarrow \nu_\mu$ and
$\bar\nu_e \rightarrow \bar\nu_\mu$ oscillation probabilities, the predicted
numbers of $\nu_\mu$ and $\bar\nu_\mu$ CC appearance events, and the
average probability for $\nu_e\to\nu_\mu$ assuming no matter effects. 
The predictions are tabulated for $\sin^22\theta=0.1$ with various 
$E_\mu$ and values of $\Delta m^2 > 0$ suggested by
the atmospheric neutrino data. 
If $\Delta m^2 < 0$ (i.e., the lower
mass eigenstate is more closely associated with $\nu_\mu$ and not
$\nu_e$), then the oscillation probabilities for neutrinos and
antineutrinos in Table~\ref{emu_tab} are reversed. Thus the sign of
$\Delta m^2$ may be determined from these measurements if matter effects
are present~\cite{bppw,lipari2}.

In this paper we have assumed that the neutrino mixing conserves $CP$;
the apparent violation of $CP$ in the $\nu_e \rightarrow \nu_\mu$ and
$\bar\nu_e \rightarrow \bar\nu_\mu$ channels would be due to matter
effects. Measuring $\nu_\mu \rightarrow \nu_e$ and $\bar\nu_\mu
\rightarrow \bar\nu_e$ oscillations as well (through an electron
appearance experiment) would provide important cross-checks and help
distinguish between matter effects and $CP$ violation. For example,
$CPT$ conservation for vacuum oscillations implies $P(\nu_e \rightarrow
\nu_\mu) = P(\bar\nu_\mu \rightarrow \bar\nu_e)$, whether $CP$ is
violated or not, so that a difference between these two oscillation
probabilities would be unambiguous evidence for matter
effects. Similarly, matter effects are the same for $\nu_e \rightarrow
\nu_\mu$ and $\nu_\mu \rightarrow \nu_e$ oscillations, so that a
difference between these oscillation channels is due to $T$ violation,
which in turn implies $CP$ violation under the plausible assumption that
$CPT$ is conserved.

\section{$\nu_{\lowercase{e}} \rightarrow \nu_\tau$ Oscillations}

If the atmospheric neutrino results are due to $\nu_\mu \rightarrow \nu_\tau$ 
oscillations with a $\Delta m^2 \sim 3.5\times10^{-3}$~eV$^2$/c$^4$, and the 
solar neutrino results are due to $\nu_e \rightarrow \nu_\mu$ and/or
$\nu_e \rightarrow \nu_\tau$ oscillations with a $\Delta m^2 \le$
O($10^{-5}$)~eV$^2$/c$^4$, it is tempting to conclude that $\nu_e
\rightarrow \nu_\tau$ oscillations should exist with an associated
$\Delta m^2 =$~O($10^{-3}$)~eV$^2$/c$^4$, but with unknown effective
$\sin^2 2 \theta$. For $\Delta m^2 > 2\times10^{-3}$~eV$^2$/c$^4$,
$\nu_e \rightarrow \nu_\tau$ oscillations with $\sin^2 2 \theta > $~0.2
are already excluded by the CHOOZ reactor $\nu_e$ disappearance
experiment, although there is no limit for $\Delta m^2 <
10^{-3}$~eV$^2$/c$^4$. In a three-neutrino scenario, measurements of
atmospheric neutrinos indicate $\sin^22\theta_{23} > 0.8$ and
exclude $\sin^2 2 \theta_{13} > $~0.33 for all atmospheric
$\Delta m^2$~\cite{bww98}; $\nu_e \rightarrow \nu_\tau$ oscillations may
be detectable if $\theta_{13}$ is not too small.
Hence, the region of parameter space that would be interesting to 
cover in future $\nu_e \rightarrow \nu_\tau$ oscillation searches is 
given by $\Delta m^2 \sim 3.5\times10^{-3}$~eV$^2$/c$^4$ 
and effective $\sin^2 2 \theta < 0.3$. 

\begin{figure}
\vspace*{.4in}
\epsfxsize4.5in
\centerline{\epsffile{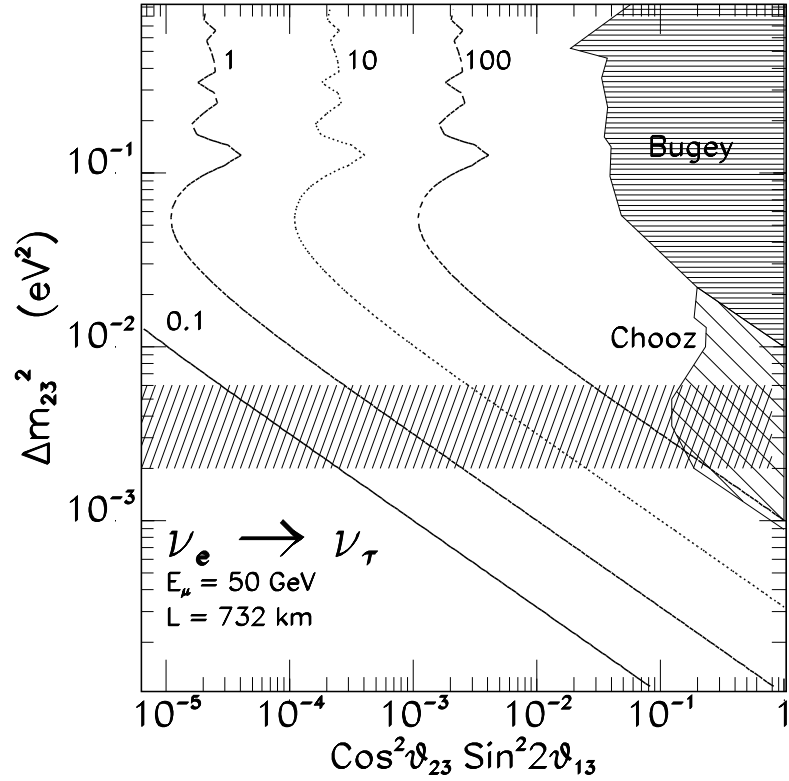}}
\vspace{1cm}
\caption{Contours of constant $\nu_\tau$ CC interaction event rate
for $\nu_e \rightarrow \nu_\tau$ appearance 732~km downstream of a
50~GeV muon storage ring neutrino source. From left to right, the
contours correspond to event rates of 0.1, 1, 10, and 100 per kt-yr.
The shaded areas are excluded by Bugey
and Chooz $\nu_e$ disappearance null results; the horizontal band
indicates the range of $\Delta m^2$ suggested by atmospheric neutrino
oscillations.
}
\label{nue_t1}
\end{figure}

\begin{figure}
\vspace*{.4in}
\epsfxsize4.5in
\centerline{\epsffile{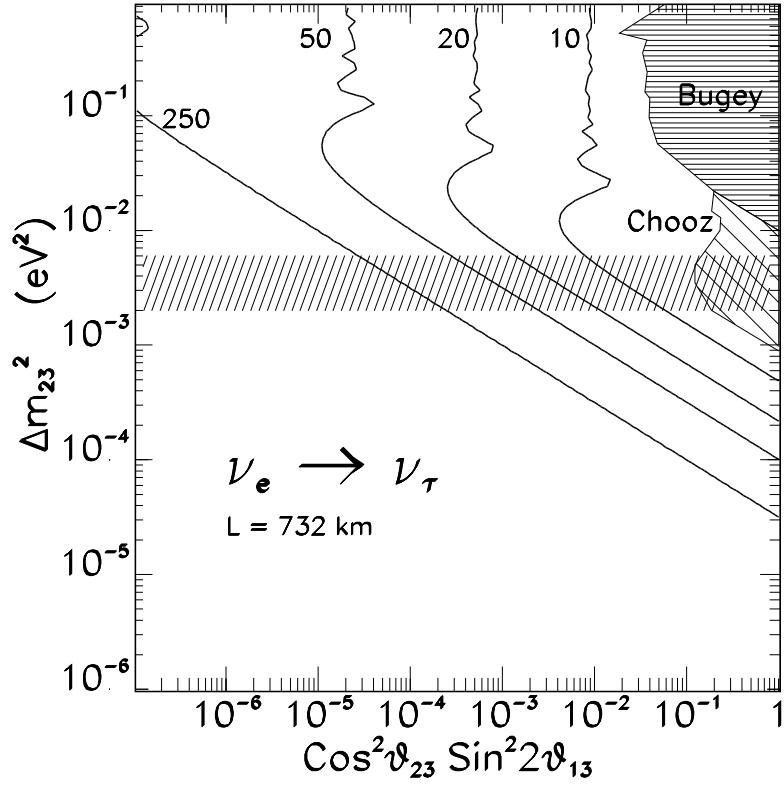}}
\vspace{1cm}
\caption{Single event per kt-yr contours calculated for 
$\nu_e \rightarrow \nu_\tau$ appearance 732~km downstream of a muon
storage ring neutrino source. From left to right, the contours
correspond to unpolarized muons with energies $E_\mu =$~250, 50, 20, and
10~GeV. The shaded areas are excluded
by Bugey and Chooz $\nu_e$ disappearance null results; the horizontal
band indicates the range of $\Delta m^2$ suggested by atmospheric
neutrino oscillations.
}
\label{nue_t2}
\end{figure}

Consider first a $\nu_e \rightarrow \nu_\tau$ appearance experiment.
Event rates are summarized in Table~\ref{etau_tab} for the neutrino
source described in Section~II.A, and $\sin^2 2 \theta = 0.1$. For a
baseline length of 732~km matter effects are small, and hence the
oscillation rates for $\sin^2 2 \theta < 0.1$ can be obtained by
multiplying the rates in the table with the value of $\sin^2 2
\theta/0.1$.  For a baseline length of 7332~km matter effects are large
and enhance the $\nu_\tau$ appearance rate for $\Delta m^2 \sim
5\times10^{-3}$~eV$^2$/c$^4$, although the numbers of events are not
large.  On the other hand, the corresponding rates for a 1~kt detector
732~km downstream yield more $\nu_\tau$ appearance events in the region
of interest, and for $E_\mu = 50$~GeV could yield up to a few events per year with $\sin^2 2\theta$ as small as $10^{-2}$ (see Fig.~\ref{nue_t1}). Single-event per
kt-yr sensitivity contours in the ($\sin^2 2 \theta$, $\Delta
m^2$)-plane are shown versus the stored muon energy in
Fig.~\ref{nue_t2}. Decreasing the storage ring energy from 50~GeV to 20~GeV would increase the minimum $\sin^2 2\theta$  that could be probed by a factor of a few.

In a three-neutrino scenario, a 1~kt detector 732~km downstream could
probe $\cos^2\theta_{23}\sin^22\theta_{13}$ to very low values; when
combined with $\nu_e \rightarrow \nu_\mu$ measurements (see Sec.~IV) and
with future atmospheric neutrino results, a precise determination of
$\theta_{23}$, $\theta_{13}$ and $\Delta m^2$ should be possible.

\section{Summary}

In this paper we have made a comprehensive study of possible neutrino
oscillation measurements in long-baseline experiments with an intense
neutrino beam from a muon storage ring. Our results can be summarized as
follows:

\begin{description}

\item{(i)} A neutrino beam from a muon storage ring provides much larger
electron neutrino interaction rates than conventional neutrino beams,
and much larger muon neutrino interaction rates if $E_\mu \ge 20$~GeV.

\item{(ii)} If the $\Delta m^2$ associated with the atmospheric
neutrino results is in the lower part of the allowed Super-Kamiokande
parameter space ($\Delta m^2 < 4\times10^{-3}$~eV$^2$/c$^4$), a Fermilab
$\rightarrow$ Gran Sasso trans-Earth
experiment with an $E_\mu = 20$~GeV muon storage ring
neutrino source would enable $\nu_\mu$ disappearance to be established
in an accelerator-based experiment, and $\Delta m^2$ to be measured
with a precision of a few percent. Matter effects may also allow the
differentiation of $\nu_\mu \rightarrow \nu_\tau$ and $\nu_\mu
\rightarrow \nu_s$ oscillations.

\item{(iii)} If the $\Delta m^2$ associated with the atmospheric neutrino
results is in the upper part of the allowed Super-Kamiokande/Kamiokande
parameter space ($\Delta m^2 > 4\times10^{-3}$~eV$^2$/c$^4$), a Fermilab
$\rightarrow$ Soudan (or CERN $\rightarrow$ Gran~Sasso) $\nu_\mu$
disappearance experiment would allow $\Delta m^2$ to be measured with a
precision of a few percent; a $\nu_\tau$ appearance experiment would
enable a search for a small contribution from $\nu_\mu
\rightarrow \nu_s$ within a dominantly $\nu_\mu \rightarrow \nu_\tau$
signal, or conversely, a small contribution from $\nu_\mu
\rightarrow \nu_\tau$ within a dominantly $\nu_\mu \rightarrow \nu_s$
signal.

\item{(iv)} If the LSND $\nu_\mu \rightarrow \nu_e$ oscillation results 
are confirmed by MiniBooNE, a Fermilab $\rightarrow$ Soudan 
$\nu_e \rightarrow \nu_\mu$ experiment would enable many thousands 
of wrong-sign muons to be measured, facilitate a precise 
confirmation and measurement of the oscillation parameters, and enable 
$P(\nu_\mu \rightarrow \nu_e)$ to be compared with 
$P(\nu_e \rightarrow \nu_\mu)$ for a test of $T$ invariance.

\item{(v)} If KamLAND observes $\nu_e$ disappearance with a rate 
corresponding to the large angle MSW solar neutrino solution then a 
measurement of $\nu_e \rightarrow \nu_\mu$ appearance might be 
concievable for a Fermilab $\rightarrow$ Soudan experiment if 
the full muon source needed for a muon collider were available 
for several years of running with a muon storage ring neutrino 
source. It might also be possible to measure matter effects as 
the electron neutrinos traverse the Earth. These measurements 
are however challenging and need further consideration to 
assess their viability.

\item{(vi)} If there is a small but non-negligible component of
$\nu_\mu \leftrightarrow \nu_e$ oscillations in atmospheric neutrinos, a
Fermilab $\rightarrow$ Gran Sasso measurement of $\nu_e \rightarrow
\nu_\mu$ and $\bar\nu_e \rightarrow \bar\nu_\mu$ could perhaps detect
matter effects in the oscillation, which show up as apparent $CP$
violation. Matter effects and true $CP$ violation can be distinguished
by also measuring $\nu_\mu \rightarrow \nu_e$ and $\bar\nu_\mu
\rightarrow \bar\nu_e$ oscillations: a nonzero $P(\nu_e \rightarrow
\nu_\mu) - P(\nu_\mu \rightarrow \nu_e)$ signals $T$ (and hence $CP$)
violation, a nonzero $P(\nu_e \rightarrow \nu_\mu) - P(\bar\nu_\mu
\rightarrow \bar\nu_e)$ signals matter effects, while a nonzero $P(\nu_e
\rightarrow \nu_\mu) - P(\bar\nu_e \rightarrow \bar\nu_\mu)$ can occur
due to either matter or $CP$ violation. If oscillations of neutrinos are
enhanced and antineutrinos suppressed, then $\Delta m^2 >0$; for the
reverse situation, $\Delta m^2 < 0$.

\item{(vii)} A $\nu_e \rightarrow \nu_\tau$ appearance search sensitive 
to $\Delta m^2$~O(10$^{-3}$)~eV$^2$/c$^4$ and $\sin^2 2\theta$ down to 
about $10^{-2}$ seems possible with a Fermilab $\rightarrow$ Soudan 
experiment using a $\ge 20$~GeV muon storage ring. Higher energies are
more desirable for $\nu_\tau$ appearance experiments.

\item{(viii)} In a three-neutrino scenario with different $\Delta m^2$
scales for the solar and atmospheric oscillations, measurements of
$\nu_e \rightarrow \nu_\mu$ and $\nu_e \rightarrow \nu_\tau$ should
provide an accurate determination of the two mixing angles associated
with the leading oscillation. When combined with solar neutrino
measurements, the three-neutrino mixing parameters would be completely
determined, except for a possible $CP$ violating phase.

\end{description}

\vspace{0.5cm}

\section*{Acknowledgments} 
The authors would like to thank D. Casper and J. Morphin for providing
tables of neutrino cross sections, E.~Kearns and J.~Learned for useful
correspondence, and J.~Hylen, S.~Parke and T.~Weiler and for discussions.
Part of this work was
performed at the Fermi National Accelerator Laboratory, which is
operated by Universities Research Association, under contract
DE-AC02-76CH03000 with the U.S. Department of Energy. This work was
supported in part by the U.S. Department of Energy, Division of High
Energy Physics, under Grants No.~DE-FG02-94ER40817, and
No.~DE-FG02-95ER40896, and in part by the University of Wisconsin
Research Committee with funds granted by the Wisconsin Alumni Research
Foundation.

\newpage

\begin{table}
\caption{\label{flux_tab}
Neutrino and antineutrino fluxes calculated for the
baseline lengths $L = 7332$~km (FNAL $\rightarrow$ Gran Sasso)
and $L = 732$~km (FNAL $\rightarrow$ Soudan).
In the calculation the neutrino fluxes have been averaged over
  a circular area with radius 1~km (0.1~km) for $L = 7332$~km (732~km). 
  This averaging provides an approximate model for the $\cal O$(10\%) reduction 
  in the neutrino fluxes at the far site due to the  
  finite divergence of the parent muon beam in the straight section 
  of the storage ring.}
\begin{center}
\begin{tabular}{c|c|cccc}
parent & $E_\mu$ (GeV) &10&20&50&250\\
\hline
& FNAL $\rightarrow$ Soudan & & & &\\
$\mu^-$ & $\nu_\mu$  ($10^{12}\;m^{-2}$ yr$^{-1}$)
&0.79&3.3&21&460\\
$\mu^-$ & $\overline{\nu_e}$  ($10^{12}\;m^{-2}$ yr$^{-1}$)
&0.79&3.3&21&470\\
$\mu^+$ & $\nu_e$  ($10^{12}\;m^{-2}$ yr$^{-1}$)
&0.85&3.3&21&470\\
$\mu^+$ & $\overline{\nu_\mu}$  ($10^{12}\;m^{-2}$ yr$^{-1}$)
&0.84&3.3&21&470\\
\hline
& FNAL $\rightarrow$ Gran Sasso & & & &\\
$\mu^-$ & $\nu_\mu$  ($10^{10}\;m^{-2}$ yr$^{-1}$)
&0.79&3.3&21&470\\
$\mu^-$ & $\overline{\nu_e}$  ($10^{10}\;m^{-2}$ yr$^{-1}$)
&0.79&3.3&21&470\\
$\mu^+$ & $\nu_e$  ($10^{10}\;m^{-2}$ yr$^{-1}$)
&0.83&3.3&21&470\\
$\mu^+$ & $\overline{\nu_\mu}$  ($10^{10}\;m^{-2}$ yr$^{-1}$)
&0.85&3.3&21&470
\end{tabular}
\end{center}
\end{table}

\begin{table}
\caption{\label{rate_tab}
Neutrino and antineutrino CC interaction rates in the
absence of oscillations, calculated for the
baseline lengths $L = 7332$~km (FNAL $\rightarrow$ Gran Sasso)
and $L = 732$~km (FNAL $\rightarrow$ Soudan).
In the calculation the neutrino fluxes have been averaged over
  a circular area with radius 1~km (0.1~km) for $L = 7332$~km (732~km).
  This averaging provides an approximate model for the $\cal O$(10\%) reduction
  in the neutrino fluxes at the far site due to the
  finite divergence of the parent muon beam in the straight section
  of the storage ring.}
\begin{center}
\begin{tabular}{c|c|cccc}
parent & $E_\mu$ (GeV) &10&20&50&250\\
\hline
& FNAL $\rightarrow$ Soudan & & & &\\
$\mu^-$ & $\nu_\mu$  (per kt-yr)
& $2.2 \times 10^3$ & $1.9 \times 10^4$
& $2.9 \times 10^5$ & $3.1 \times 10^7$\\
$\mu^-$ & $\overline{\nu_e}$  (per kt-yr)
& $9.6 \times 10^2$ & $8.0 \times 10^3$
& $1.3 \times 10^5$ & $1.4 \times 10^7$\\
$\mu^+$ & $\nu_e$  (per kt-yr)
& $1.9 \times 10^3$ & $1.6 \times 10^4$
& $2.4 \times 10^5$ & $2.7 \times 10^7$\\
$\mu^+$ & $\overline{\nu_\mu}$  (per kt-yr)
& $1.2 \times 10^3$ & $9.5 \times 10^3$
& $1.5 \times 10^5$ & $1.6 \times 10^7$\\
\hline
& FNAL $\rightarrow$ Gran Sasso & & & &\\
$\mu^-$ & $\nu_\mu$  (per kt-yr)
& 22 & $1.9 \times 10^2$ & $2.9 \times 10^3$ &$3.1 \times 10^5$\\
$\mu^-$ & $\overline{\nu_e}$  (per kt-yr)
& 9.5 & 80 & $1.3 \times 10^3$ & $1.4 \times 10^5$\\
$\mu^+$ & $\nu_e$  (per kt-yr)
& 20 & $1.6 \times 10^2$ & $2.4 \times 10^3$  & $2.7 \times 10^5$\\
$\mu^+$ & $\overline{\nu_\mu}$  (per kt-yr)
& 12 & 94 & $1.5 \times 10^3$ & $1.6 \times 10^5$
\end{tabular}
\end{center}
\end{table}

\begin{table}
\caption{\label{compare_tab}
Muon neutrino and electron antineutrino CC interaction rates in the
absence of oscillations, calculated for baseline length $L = 732$~km
(FNAL $\rightarrow$ Soudan), for MINOS using the wide band beam and a
muon storage ring with $E_\mu=10, 20, 50$ and $250$~GeV.
}
\begin{center}
\begin{tabular}{cc|cc|cc}
& & $\langle E_{\nu_\mu} \rangle$ & $\langle E_{\bar \nu_e} \rangle$
& N($\nu_\mu$ CC) & N($\bar\nu_e$ CC) \\
Experiment & & (GeV) & (GeV) & (per kt-yr) & (per kt-yr) \\
\hline
MINOS & Beam & & & & \\
\hline
& Low energy    &  3 & -- &  458 & 1.3 \\
& Medium energy &  6 & -- & 1439 & 0.9 \\
& High energy   & 12 & -- & 3207 & 0.9 \\
\hline
Muon ring & $E_\mu$ (GeV) & & & & \\
\hline
&  10 &   7 &   6 & 2200              & 960 \\
&  20 &  14 &  12 & 19000             & 8000 \\
&  50 &  35 &  30 & 2.9$\times$10$^5$ & 1.3$\times$10$^5$ \\
& 250 & 175 & 150 & 3.1$\times$10$^7$ & 1.4$\times$10$^7$
\end{tabular}
\end{center}
\end{table}

\begin{table}
\caption{\label{densities_tab}
Tilt angle, distance, and average electron density for
long-baseline experiments with neutrino source at Fermilab.
}
\begin{center}
\begin{tabular}{c|ccc|ccc} 
Detector & Tilt angle & $L$  & $\langle N_e \rangle$
& $L_{core}$ & $\langle N_e \rangle_{mantle}$ & $\langle N_e \rangle_{core}$\\
site     & (degrees)  & (km) & ($N_0$/cm$^3$)
& (km) & ($N_0$/cm$^3$) & ($N_0$/cm$^3$) \\
\hline
Soudan     &  3.3 &   732 & 1.67 &    0 & 1.67 & -- \\
Gran Sasso & 35   &  7332 & 2.09 &    0 & 2.09 & -- \\
Kamioka    & 46   &  9160 & 2.26 &    0 & 2.26 & -- \\
South Pole & 67   & 11700 & 3.39 & 4640 & 2.33 & 4.99 
\end{tabular}
\end{center}
\end{table}

\begin{table}
\caption[]{\label{osc_rate_tab}
Muon neutrino CC interaction rates calculated for a distance $L$ downstream of a storage ring containing unpolarized negative muons. The event rates assume $\nu_\mu \rightarrow \nu_\tau$ oscillations with $\sin^2 2\theta = 1$. Results for $\nu_\mu \rightarrow \nu_s$
oscillations are shown in parentheses for $L=7332$~km; for $L=732$~km
matter effects are small and the results for $\nu_\mu \rightarrow
\nu_\tau$ and $\nu_\mu \rightarrow \nu_s$ are very similar. Also shown
are the tau neutrino CC interaction rates due to $\nu_\mu \rightarrow
\nu_\tau$ oscillations.
}
\begin{center}
\begin{tabular}{ccc|ccc|c}
$E_\mu$& $L$ & $\Delta m^2$&$\left<P\right>$ &N($\nu_\mu$ CC) &Significance& N($\nu_\tau$ CC) \\
(GeV)& (km) &(eV$^2$/c$^4$)& $\nu_\mu \rightarrow \nu_\tau$
($\nu_\mu \rightarrow \nu_s$) &(per 10 kt-yr)& & (per kt-yr)\\
\hline
10 & 7332 & 0      & 0 (0)       & 220 (220) & 0 (0) & 0 \\
10 & 7332 & 0.002  & 0.53 (0.18) & 100 (180) & $12\sigma$ ($3.0\sigma$) & 1.9 \\
10 & 7332 & 0.003  & 0.29 (0.36) & 160 (140) & $5.2\sigma$ ($6.7\sigma$) & 0.7 \\
10 & 7332 & 0.004  & 0.63 (0.64) &  80 ( 80) & $15\sigma$ ($16\sigma$)  & 2.1 \\
10 & 7332 & 0.005  & 0.56 (0.48) & 100 (120) & $12\sigma$ ($9.9\sigma$) & 1.9 \\
10 & 7332 & 0.006  & 0.38 (0.37) & 140 (140) & $7.1\sigma$($7.0\sigma$) & 1.2 \\
\hline
20 & 7332 & 0     & 0 (0)   & 1900 (1900) & 0 (0) & 0 \\
20 & 7332 & 0.002 & 0.80 (0.27) &  370 (1370) & $78\sigma$  ($14\sigma$) & 43 \\
20 & 7332 & 0.003 & 0.81 (0.25) &  360 (1400) & $80\sigma$  ($12\sigma$) & 46 \\
20 & 7332 & 0.004 & 0.53 (0.18) &  880 (1530) & $34\sigma$  ($8.6\sigma$) & 30 \\
20 & 7332 & 0.005 & 0.30 (0.21) & 1310 (1480) & $15\sigma$  ($10\sigma$) & 15 \\
20 & 7332 & 0.006 & 0.29 (0.36) & 1320 (1190) & $15\sigma$  ($20\sigma$) & 14 \\
\hline
50 & 7332 & 0     & 0 (0)   & 28800 (28800) & 0 (0) & 0 \\
50 & 7332 & 0.002 & 0.26 (0.10)   & 21300 (25900) & 52$\sigma$ (18$\sigma$) & 350 \\
50 & 7332 & 0.003 & 0.48 (0.18)   & 15000 (23700) & 110$\sigma$ (33$\sigma$) & 660 \\
50 & 7332 & 0.004 & 0.67 (0.24)   & 9500 (22000) & 200$\sigma$ (46$\sigma$) & 950 \\
50 & 7332 & 0.005 & 0.80 (0.27)   & 5700 (21000) & 310$\sigma$ (54$\sigma$) & 1160 \\
50 & 7332 & 0.006 & 0.86 (0.28)   & 4200 (20800) & 380$\sigma$ (55$\sigma$) & 1260 \\
\hline
\hline
10 & 732 & 0     & 0    & 22200 & 0 & 0 \\
10 & 732 & 0.002 & 0.08 & 20500 &  $12\sigma$ & 20 \\
10 & 732 & 0.003 & 0.16 & 18700 &  $26\sigma$ & 43 \\
10 & 732 & 0.004 & 0.26 & 16500 &  $45\sigma$ & 74 \\
10 & 732 & 0.005 & 0.37 & 14100 &  $69\sigma$ & 110 \\
10 & 732 & 0.006 & 0.47 & 11700 &  $98\sigma$ & 150 \\
\hline
20 & 732 & 0     & 0     & 185000 & 0 & 0 \\
20 & 732 & 0.002 & 0.020 & 181000 &  $8.6\sigma$ & 89 \\
20 & 732 & 0.003 & 0.044 & 177000 &  $19\sigma$ & 200 \\
20 & 732 & 0.004 & 0.076 & 171000 &  $34\sigma$ & 350 \\
20 & 732 & 0.005 & 0.11  & 164000 &  $52\sigma$ & 530 \\
20 & 732 & 0.006 & 0.16  & 156000 &  $74\sigma$ & 750 \\
\hline
50 & 732 & 0     & 0     &$2.88 \times 10^6$& 0 & 0 \\
50 & 732 & 0.002 &$3.3 \times 10^{-3}$&$2.88 \times 10^6$&  5.6$\sigma$ & 410 \\
50 & 732 & 0.003 &$7.4 \times 10^{-3}$&$2.86 \times 10^6$&  13$\sigma$ & 910 \\
50 & 732 & 0.004 &0.013 &$2.84 \times 10^6$&  22$\sigma$ & 1600 \\
50 & 732 & 0.005 &0.020&$2.83 \times 10^6$& 35$\sigma$ & 2500 \\
50 & 732 & 0.006 &0.029&$2.81 \times 10^6$& 50$\sigma$ & 3600 \\
\hline
250 & 732 & 0     & 0     &$3.11 \times 10^8$& 0 & 0 \\
250 & 732 & 0.002 &$1.4 \times 10^{-4}$&$3.11 \times 10^8$&  2.5$\sigma$ & 3000 \\
250 & 732 & 0.003 &$3.2 \times 10^{-4}$&$3.11 \times 10^8$&  5.7$\sigma$ & 6800 \\
250 & 732 & 0.004 &$5.8 \times 10^{-4}$&$3.11 \times 10^8$&  10$\sigma$ & 12100 \\
250 & 732 & 0.005 &$9.0 \times 10^{-4}$&$3.11 \times 10^8$& 16$\sigma$ & 18900 \\
250 & 732 & 0.006 &$1.3 \times 10^{-3}$&$3.11 \times 10^8$& 23$\sigma$ & 27200 \end{tabular}
\end{center}
\end{table}


\begin{table}
\caption[]{\label{emu_tab}
Average $\nu_e \rightarrow \nu_\mu$ and $\bar\nu_e \rightarrow
\bar\nu_\mu$ oscillation probabilities, and muon neutrino and antineutrino
CC interaction rates for a detector $L=7332$~km downstream of a storage ring containing unpolarized muons. The event rates assume $\nu_e \rightarrow
\nu_\mu$ oscillations with $\sin^2 2\theta = 0.1$. Also shown is the
average oscillation probability for $\nu_e \rightarrow \nu_\mu$,
assuming no matter effects; the probability for $\bar\nu_e \rightarrow
\bar\nu_\mu$ without matter effects is the same as for $\nu_e
\rightarrow \nu_\mu$.
}
\begin{center}
\begin{tabular}{cc|cc|cc|c}
$E_\mu$ & $\Delta m^2$ & & N($\nu_\mu$ CC) & & N($\bar\nu_\mu$ CC)
& No matter effects\\
(GeV) & (eV$^2$/c$^4$)
& $\left<P(\nu_e \rightarrow \nu_\mu)\right>$ &(per 10 kt-yr)
& $\left<P(\bar\nu_e \rightarrow \bar\nu_\mu)\right>$ & (per 10 kt-yr)
& $\left<P(\nu_e \rightarrow \nu_\mu)\right>$ \\
\hline
10 & 0.002  & 0.46   &   90 & 0.015  & 1.4 & 0.041 \\
10 & 0.003  & 0.53   &  110 & 0.019  & 1.8 & 0.042 \\
10 & 0.004  & 0.27   &   55 & 0.032  & 3.0 & 0.061 \\
10 & 0.005  & 0.10   &   20 & 0.031  & 2.9 & 0.047 \\
10 & 0.006  & 0.087  &   18 & 0.030  & 2.9 & 0.046 \\
\hline
20 & 0.002  & 0.11   &  170 & 0.011  & 8.8 & 0.082 \\
20 & 0.003  & 0.29   &  460 & 0.016  & 13  & 0.069 \\
20 & 0.004  & 0.46   &  730 & 0.015  & 12  & 0.041 \\
20 & 0.005  & 0.55   &  870 & 0.014  & 11  & 0.032 \\
20 & 0.006  & 0.53   &  840 & 0.019  & 15  & 0.042 \\
\hline
50 & 0.002  & 0.0067 &  170 & 0.0016 &  21 & 0.033 \\
50 & 0.003  & 0.025  &  630 & 0.0044 &  57 & 0.057 \\
50 & 0.004  & 0.059  & 1480 & 0.0079 & 110 & 0.074 \\
50 & 0.005  & 0.11   & 2750 & 0.011  & 140 & 0.082 \\
50 & 0.006  & 0.18   & 4380 & 0.014  & 180 & 0.082
\end{tabular}
\end{center}
\end{table}

\clearpage

\begin{table}
\caption[]{\label{etau_tab}
Electron neutrino CC interaction rates calculated for a distance $L$
downstream of a storage ring containing unpolarized positive muons
assuming $\nu_e \rightarrow \nu_\tau$ oscillations with $\sin^2 2\theta
= 0.1$. Also shown are the tau neutrino CC interaction rates due to
oscillations.
}
\begin{center}
\begin{tabular}{ccc|ccc|c}
$E_\mu$& $L$ & $\Delta m^2$&$\left<P\right>$ &N($\nu_e$ CC) &Significance& N($\nu_\tau$ CC)
\\
(GeV)& (km) &(eV$^2$/c$^4$)& &(per 10 kt-yr)& & (per kt-yr)\\
\hline
10 & 7332 & 0      & 0     & 200 & 0           & 0 \\
10 & 7332 & 0.002  & 0.46  & 110 & $8.9\sigma$ & 1.1 \\
10 & 7332 & 0.003  & 0.53  &  95 & $ 11\sigma$ & 1.6 \\
10 & 7332 & 0.004  & 0.27  & 150 & $4.5\sigma$ & 0.9 \\
10 & 7332 & 0.005  & 0.10  & 180 & $1.5\sigma$ & 0.3 \\
10 & 7332 & 0.006  & 0.087 & 180 & $1.3\sigma$ & 0.2 \\
\hline
20 & 7332 & 0      & 0      & 1580 & 0           & 0 \\
20 & 7332 & 0.002  & 0.11   & 1410 & $4.5\sigma$ & 7.2 \\
20 & 7332 & 0.003  & 0.29   & 1120 & $ 14\sigma$ & 14 \\
20 & 7332 & 0.004  & 0.46   &  850 & $ 25\sigma$ & 17 \\
20 & 7332 & 0.005  & 0.55   &  710 & $ 33\sigma$ & 16 \\
20 & 7332 & 0.006  & 0.53   &  740 & $ 31\sigma$ & 16 \\
\hline
50 & 7332 & 0      & 0      & 24100 & 0           & 0 \\
50 & 7332 & 0.002  & 0.007 & 24000 & 1.0$\sigma$   & 4.2 \\
50 & 7332 & 0.003  & 0.026 & 23000 & 3.9$\sigma$   & 19 \\
50 & 7332 & 0.004  & 0.062 & 22100 & 9.8$\sigma$   & 51 \\
50 & 7332 & 0.005  & 0.12  & 20900 & 19$\sigma$   & 100 \\
50 & 7332 & 0.006  & 0.18  & 19300 & 31$\sigma$   & 170 \\
\hline
\hline
10 & 732 & 0      & 0      & 19400 & 0           & 0 \\
10 & 732 & 0.002  & 0.010  & 19200 & $1.4\sigma$ & 1.7 \\
10 & 732 & 0.003  & 0.021  & 19000 & $3.0\sigma$ & 3.7 \\
10 & 732 & 0.004  & 0.033  & 18800 & $4.7\sigma$ & 6.3 \\
10 & 732 & 0.005  & 0.045  & 18500 & $6.4\sigma$ & 9.2 \\
10 & 732 & 0.006  & 0.057  & 18300 & $8.2\sigma$ & 12 \\
\hline
20 & 732 & 0      & 0      & 156000 & 0           & 0 \\
20 & 732 & 0.002  & 0.0028 & 156000 & $1.1\sigma$ & 8.8 \\
20 & 732 & 0.003  & 0.0060 & 155000 & $2.4\sigma$ & 20 \\
20 & 732 & 0.004  & 0.010  & 154000 & $4.0\sigma$ & 34 \\
20 & 732 & 0.005  & 0.015  & 154000 & $6.0\sigma$ & 52 \\
20 & 732 & 0.006  & 0.021  & 153000 & $8.3\sigma$ & 73 \\
\hline
50 & 732 & 0      & 0      & $2.37 \times 10^6$ & 0           & 0 \\
50 & 732 & 0.002  &$5 \times 10^{-4}$&$2.37 \times 10^6$&0.7$\sigma$&42 \\
50 & 732 & 0.003  &$1 \times 10^{-3}$&$2.37 \times 10^6$&1.6$\sigma$&95 \\
50 & 732 & 0.004  &$2 \times 10^{-3}$&$2.37 \times 10^6$&2.8$\sigma$&170 \\
50 & 732 & 0.005  &$3 \times 10^{-3}$&$2.36 \times 10^6$&4.4$\sigma$&260 \\
50 & 732 & 0.006  &$4 \times 10^{-3}$&$2.36 \times 10^6$&6.3$\sigma$&380 \\
\hline
250 & 732 & 0      & 0      &$2.73 \times 10^8$& 0           & 0 \\
250 & 732 & 0.002  &$2 \times 10^{-5}$&$2.73 \times 10^8$&0.3$\sigma$&360 \\
250 & 732 & 0.003  &$5 \times 10^{-5}$&$2.73 \times 10^8$&0.8$\sigma$&810 \\
250 & 732 & 0.004  &$8 \times 10^{-5}$&$2.73 \times 10^8$&1.3$\sigma$&1400 \\
250 & 732 & 0.005  &$1 \times 10^{-4}$&$2.73 \times 10^8$&2.1$\sigma$&2300 \\
250 & 732 & 0.006  &$2 \times 10^{-4}$&$2.73 \times 10^8$&3.0$\sigma$&3200 
\end{tabular}
\end{center}
\end{table}

\clearpage

\end{document}